\documentclass[12pt]{article}
\usepackage[paper=a4paper,margin=1in]{geometry}
\usepackage{t1enc}      

\usepackage{amsmath}
\usepackage{amsfonts}
\usepackage{amssymb}
\usepackage{amsthm}
\usepackage{graphicx}
\usepackage{mathrsfs}  

\newcommand{\edth} {\mbox{\symbol{'360}}}

\newcommand{\ua}{\underline a \,}

\newcommand{\uA}{\underline A \,}
\newcommand{\uB}{\underline B \,}

\numberwithin{equation}{section}

\begin{document}
\bibliographystyle{unsrt}

\title{A review of total energy-momenta in GR with a positive cosmological
constant\footnote{Dedicated to our friend, J\"org Frauendiener, on the
occasion of his 60th birthday.
\newline\newline\emph{Ach, wie sch\"on muss sich's ergehen,
\newline Dort in Newmans Himmelsraum.
\newline Und Lichtkugeln auf den H\"ohen --
\newline O wie lab'n sie meinen Traum !}
\newline
\newline
-- \emph{from} Ein theoretisch--physikalischer Schiller--Traum \emph{by}
N. V. Mitskievitch}}
\author{L\'aszl\'o B. Szabados \\
Wigner Research Centre for Physics, \\
H-1525 Budapest 114, P. O. Box 49, European Union \\
Paul Tod \\
Mathematical Institute, Oxford University, \\
Oxford OX2 6GG}

\maketitle

\begin{abstract}
A review is given of the various approaches to and expressions for total
energy-momentum and mass in the presence of a positive cosmological constant
in Einstein's field equations, together with a discussion of the key
conceptual questions, main ideas and techniques behind them.
\end{abstract}


\section{Introduction}
\label{sec-1}

As a consequence of the equivalence principle, following the lesson of the
Galileo--E\"otv\"os experiments, gravity is known to have a universal nature
and the spacetime metric plays a double role: both as the field variable for
gravity and as the geometric background for the dynamics of the matter fields
at the same time. A direct consequence of the lack of any non-dynamical
geometric background is that any local expression for the gravitational
energy-momentum is \emph{necessarily} pseudotensorial or, in the tetrad
formulation of the theory, Lorentz gauge dependent (see e.g.
\cite{Trautman,MTW,Goldberg}). Although all these local quantities can be
recovered from (various forms of) a single geometric object \emph{on the
linear frame bundle} (or of its subbundles) (see e.g.
\cite{Sparl82,DuMa,Joerg89,Sz92}), the gravitational energy-momentum density
\emph{in the spacetime} is not well defined -- it cannot be localized to
points. The gravitational energy-momentum is \emph{non-local} in nature, and
can be well defined only if it is associated to \emph{extended} domains
in spacetime. Thus, it must be a measure of \emph{total} energy-momentum,
i.e. when the domain is an infinitely large part of the spacetime (for
example, the whole spacetime itself), or \emph{quasi-local}, when the domain
is only a compact subset of the spacetime.

From pragmatic points of view the significance of these quantities is given
by the positivity of the corresponding total (or quasi-local) energy/mass,
or at least their boundedness from below. In fact, some of these quantities
have already been proven to be useful tools in geometric analysis (e.g.
of asymptotic structure of spacetime or black-hole uniqueness), in stability
investigations of general relativistic gravitating systems, in numerical
calculations (e.g. to control errors), and so on.

The usual form of the total energy-momentum of a \emph{localized gravitating
system} is a 2-surface integral of some local `superpotential' $U$, which also
depends (linearly) on some vector field $K^a$ or (quadratically) on some
spinor field $\lambda^A$ representing `asymptotic translations' in the
spacetime:

\begin{equation}
{\tt P}_{\ua}K^{\ua}:=\oint_{\cal S}U(K){\rm d}{\cal S}, \hskip 20pt {\rm or}
\hskip 20pt
{\tt P}_{\ua}\sigma^{\ua}_{\uA{\uB}'}\lambda^{\uA}\bar\lambda^{{\uB}'}:=\oint
_{\cal S}U(\lambda,\bar\lambda){\rm d}{\cal S}. \label{eq:1.1}
\end{equation}
Here $K^{\ua}$ and $\lambda^{\uA}$ are the \emph{components} of the `asymptotic 
translation' $K^a$ and its spinor constituent $\lambda^A$, respectively, in 
some basis of the \emph{space of (candidate) asymptotic 
translations}\footnote{As we shall see, this can be a problematic notion with 
$\Lambda\not=0$.}, and $\sigma^{\ua}_{\uA{\uB}'}$, ${\ua}=0,...,3$, ${\uA},{\uB}
=0,1$, are the standard $SL(2,\mathbb{C})$ Pauli matrices. Thus, the 
energy-momentum 4-vector is an element of the \emph{dual space} of the space 
of asymptotic translations. Therefore, to have a well defined total 
energy-momentum expression, we should specify: (1) the domain of integration 
${\cal S}$ (i.e. the choice of what to consider as the physical system); (2) 
the `superpotential' $U$; and (3) the `generator' ($K^a$ or $\lambda^A$) of 
the quantity in question (i.e. the \emph{definition} of the asymptotic 
translations). Different choices for these yield different expressions with 
different properties.

\emph{Total mass} (rather than energy-momentum) can be associated not only
with localized sources, but with \emph{closed universes} with non-negative
$\Lambda$ as well. In this case the domain of integration is a Cauchy surface
(rather than a closed 2-surface).

The aim of the present paper is to review the \emph{total}
energy-momentum/mass constructions in the presence of a strictly positive
cosmological constant $\Lambda$. However, to put them in perspective and to
see the roots of certain key ideas behind the actual constructions, and also
to motivate how to choose the domain, the superpotential and the generator
field in items (1)-(3) above, we briefly recall the analogous constructions
in the $\Lambda=0$ and $\Lambda<0$ cases where these ideas appeared first.
This part of the review is far from being complete. For a review of the
quasi-local constructions, see e.g. \cite{SzRev}.

\subsection{On total energy-momenta with $\Lambda=0$}
\label{sub-1.1}

\subsubsection{The domain of integration}
\label{sub-1.1.1}

\begin{figure}[ht]
\begin{center}
\includegraphics[width=0.3\textwidth]{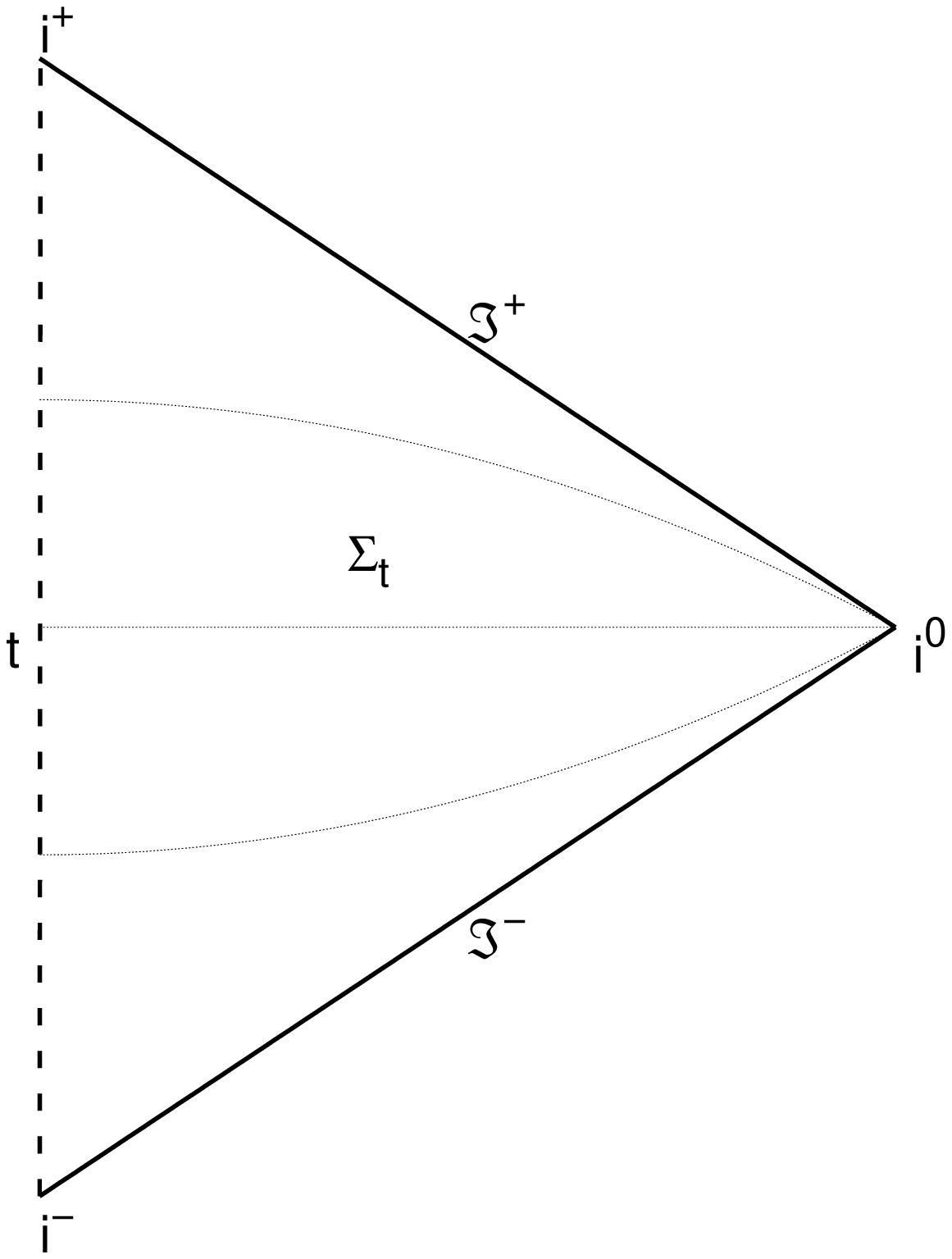}
\includegraphics[width=0.3\textwidth]{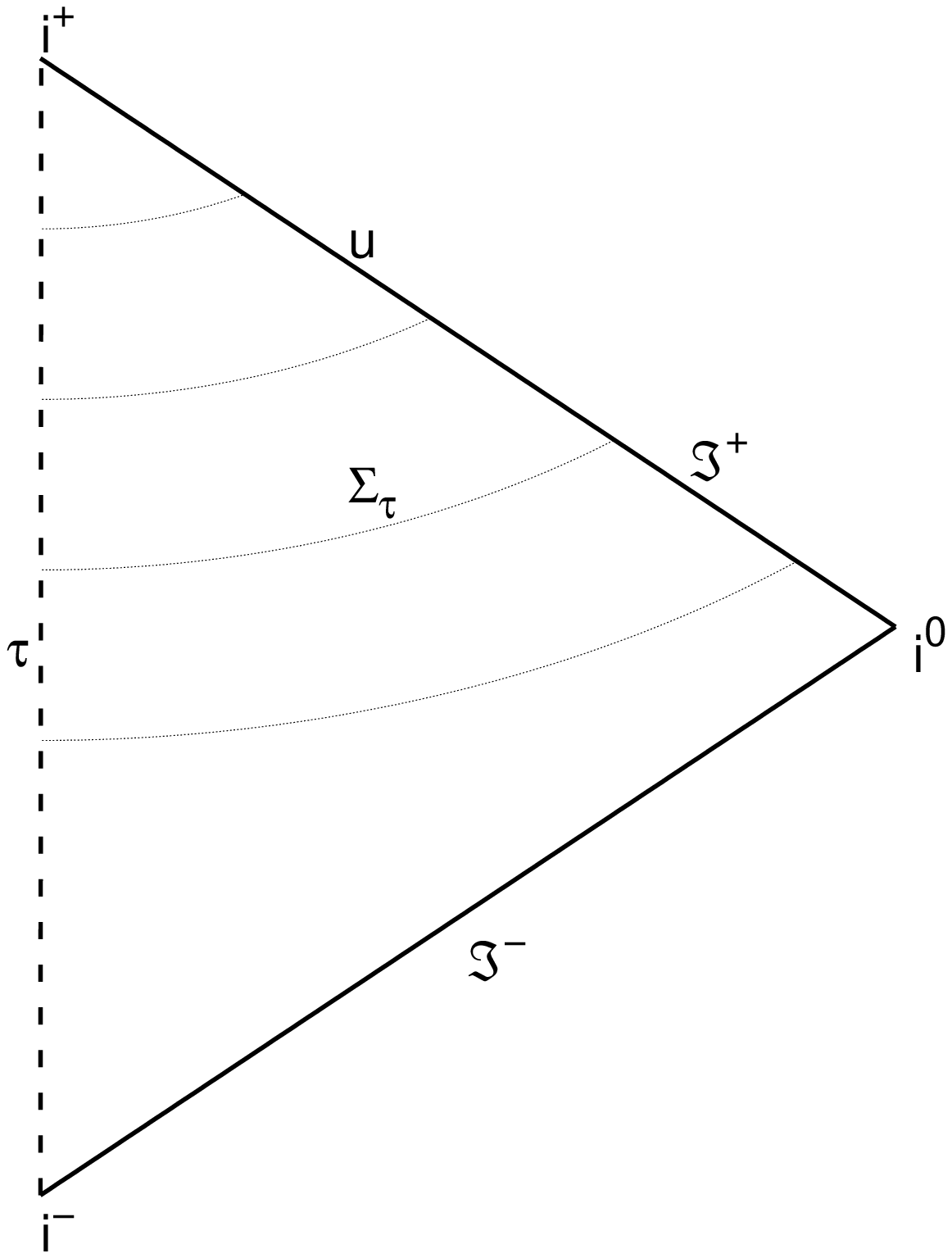}
\includegraphics[width=0.3\textwidth]{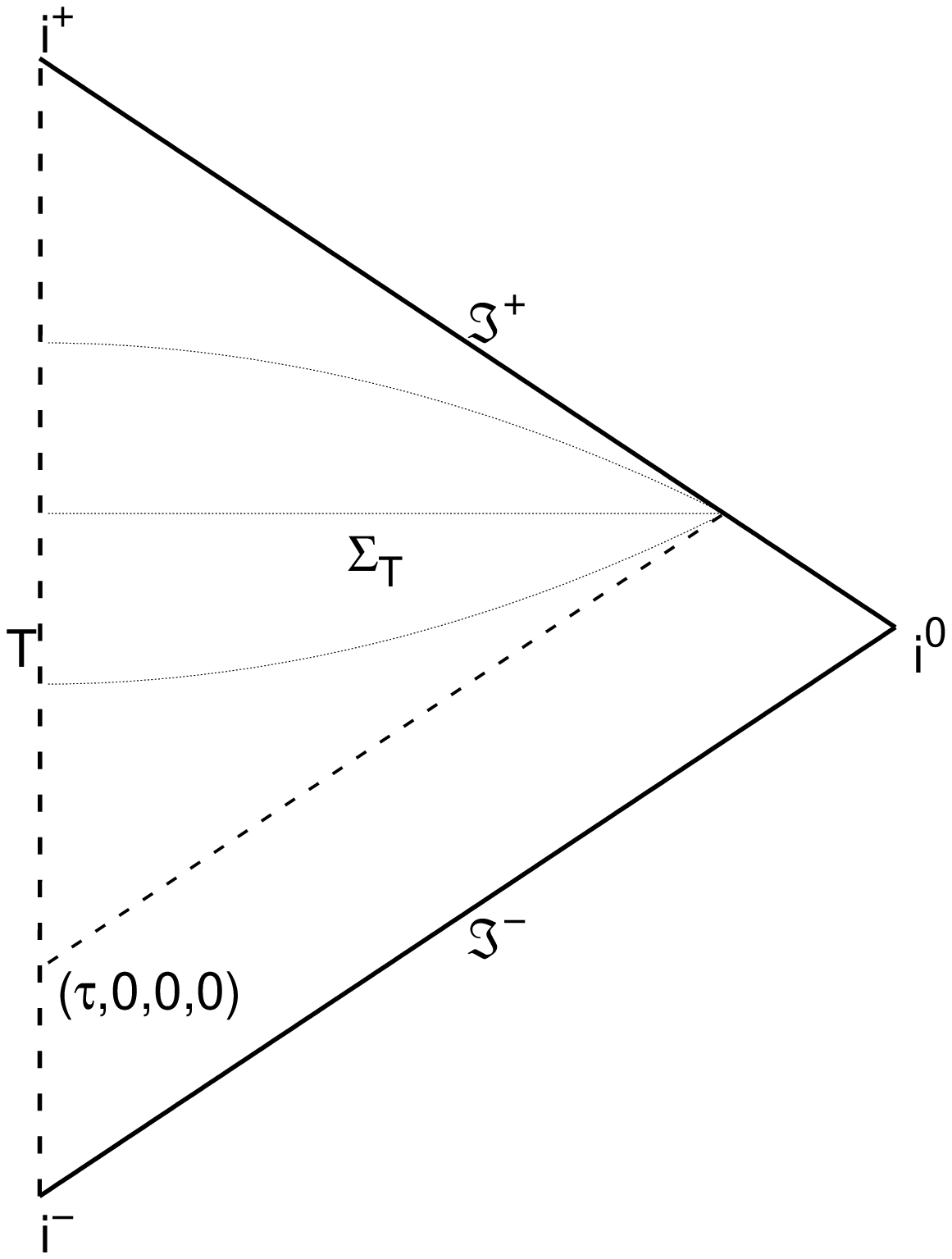}
\caption{\label{fig:Minkowski}
Foliations of the Minkowski spacetime: i. The hypersurfaces $\Sigma_t$ form a 
global foliation. Each of the leaves is a global Cauchy surface with 
$\mathbb{R}^3$ topology, and all these extend to spatial infinity $i^0$. These 
hypersurfaces are both intrinsically and extrinsically asymptotically flat. 
The  $t={\rm const}$ hyperplanes in the Cartesian coordinates are like this. 
ii. The hypersurfaces $\Sigma_\tau$, which are only \emph{partial} Cauchy 
surfaces, extend to the future null infinity $\mathscr{I}^+$ and foliate both 
the spacetime and $\mathscr{I}^+$. These hypersurfaces are intrinsically 
asymptotically hyperboloidal and the extrinsic curvature is asymptotically 
proportional to the intrinsic metric. For fixed $T>0$, in the Cartesian 
coordinates, the hypersurfaces $\tau:=t-\sqrt{T^2+r^2}={\rm const}$ have this 
character. 
iii. However, for fixed $\tau$ and variable $T$, the hypersurfaces $\Sigma
_T$, given by $T^2:=(t-\tau)^2-r^2={\rm const}$, foliate only the chronological 
future of the point $(\tau,0,0,0)$ but do not provide a foliation of (even an 
open subset of) $\mathscr{I}^+$. }
\end{center}
\end{figure}

If the cosmological constant is zero, then the spacetime describing the 
gravitational `field' of a \emph{localized} source is expected to be 
`asymptotically flat' in some well-defined sense, and hence its global 
structure at large distances is expected to be similar to that of the 
Minkowski spacetime. (For the global properties of the latter, see e.g. 
\cite{HE}.) In fact, formally the \emph{conserved} Arnowitt--Deser--Misner 
(ADM) energy-momentum \cite{ADM} is based on a 2-surface integral on the 
boundary at infinity of an (appropriately defined) asymptotically flat 
spacelike hypersurface that extends to \emph{spatial infinity}. In Minkowski 
spacetime the $t={\rm const}$ hyperplanes, denoted by $\Sigma_t$, provide a 
foliation of the spacetime by such global Cauchy surfaces, while their 
(common) boundary at infinity is the $r\rightarrow\infty$ limit of the 
2-spheres ${\cal S}_{t,r}:=\{t={\rm const}\, ,\, r={\rm const}\,\}$ (see 
Fig.~\ref{fig:Minkowski}.i.). Clearly, this boundary is \emph{spatially 
separated} from the world tube of the source. 

On the other hand, if the localized system emits radiation and the total 
energy-momentum is expected to reflect the dynamical aspects of the system, 
e.g. to give account of the energy carried away by the radiation, then that 
should be associated with a hypersurface extending to the \emph{null 
infinity} of the spacetime. In fact, the Trautman--Bondi--Sachs (TBS) 
energy-momentum \cite{Trautman58,Bondietal,Sachs} is an integral on the 
boundary at infinity of an outgoing \emph{null hypersurface} that defines a 
retarded time instant. One may also think of this 2-surface as the boundary 
at infinity of an \emph{asymptotically hyperboloidal} spacelike hypersurface 
whose asymptote is just the null hypersurface. For example, in Minkowski 
spacetime, the family of spacelike hypersurfaces $\tau:=t-\sqrt{T^2+r^2}=
{\rm const}$ for some constant $T>0$, denoted by $\Sigma_\tau$, provides 
a foliation of the whole spacetime (see Fig.~\ref{fig:Minkowski}.ii.). 
$\Sigma_\tau$ intersects future null infinity at the retarded time instant 
$u=\tau$. The induced metric $h_{ab}$ on them is of constant curvature with 
the scalar curvature ${\cal R}=-6/T^2$, and the extrinsic curvature is $\chi
_{ab}=h_{ab}/T$. Hence, these are \emph{hyperboloidal}, but only \emph{partial} 
Cauchy surfaces\footnote{It 
might be interesting to note that the line element of the flat metric with 
the global foliations $\Sigma_t$ and $\Sigma_\tau$ takes the 
Friedman--Robertson--Walker (FRW) form with $k=0$ and $k=-1$, respectively. 
On the other hand, for fixed $\tau$ and variable $T$ the hypersurfaces 
$\Sigma_T:=\{T={\rm const}\}$ foliate only the chronological future of the 
point with Cartesian coordinates $(\tau,0,0,0)$. All the leaves of this 
latter foliation intersect the future null infinity at one and the same cut 
$u=\tau$, rather than providing a foliation of the null infinity (see 
Fig.~\ref{fig:Minkowski}.iii.).}. The $r\rightarrow\infty$ limit of the 
2-spheres ${\cal S}_{\tau,r}:=\{\tau={\rm const}\, ,\, r={\rm const}\,\}$ is 
\emph{null separated} from the source in the centre $r=0$ at $t=\tau$. This 
energy-momentum depends on the retarded time instant $u$ that the 
hypersurface defines, and, in particular, the mass, defined to be the 
Lorentzian length of the TBS energy-momentum 4-vector, is a monotonically 
\emph{decreasing} function of $u$ (`Bondi's mass-loss'). To derive this, a 
foliation of the null infinity and the asymptotic solutions of the field 
equations are needed.

In the \emph{conformal approach} of Penrose \cite{Pe65,PR2} the future null 
infinity, as a smooth boundary $\mathscr{I}^+$, is attached to the conformal 
spacetime. In this framework the asymptotic flatness is \emph{defined} by the 
compactifiability of the spacetime by such a boundary. Since $\Lambda=0$,
$\mathscr{I}^+$ is a \emph{null} hypersurface in the conformal spacetime, 
its null geodesic generators are shear-free (and, in a certain conformal 
gauge, divergence-free, too); and it is assumed to have $S^2\times\mathbb{R}$ 
topology\footnote{This can be proved, given certain causality assumptions; 
see \cite{Newm}.}. The advantage of this approach to asymptotic flatness is 
that the techniques of local differential geometry can be used to study the 
asymptotic properties of the fields and spacetime (see also 
\cite{Ge,NT80,PR2,JoergLivRev}). In particular, the $r\rightarrow\infty$ 
limit of the large spheres ${\cal S}_{\tau,r}$ is the regular 2-surface 
$\Sigma_\tau\cap\mathscr{I}^+$, a so-called cut, of the null infinity 
$\mathscr{I}^+$. The difference of the energy-momenta on two different cuts 
can be written as a flux integral on the piece ${\cal I}\subset\mathscr{I}^+$ 
between the two cuts \cite{GeWi}, and its vanishing, i.e. the absence of 
outgoing gravitational radiation, is equivalent to the vanishing of the 
\emph{news function} \cite{Bondietal}  or equivalently of the magnetic part 
of the rescaled Weyl curvature, defined by the null normal of $\mathscr{I}^+$, 
on ${\cal I}$ \cite{ABK1}. 

Nevertheless, the existence of null infinity as a \emph{smooth} boundary of 
the conformal spacetime restricts the fall-off rate of the physical metric: 
it should tend to the flat metric as $1/r$. For a treatment of asymptotically 
flat spacetimes with slower (e.g. logarithmic) fall-off metrics, see. e.g. 
\cite{slowfalloff}.

\subsubsection{The asymptotic translations}
\label{sub-1.1.2}

In Minkowski spacetime the four independent translational Killing fields are
defined in a geometric/algebraic way: they are \emph{constant vector fields};
and, also, they form a \emph{commutative ideal} in the Poincar\'e algebra of
the Killing fields. At spatial infinity their restriction to the boundary
2-surface at infinity yields \emph{constant} vector fields there. Since the
isometries of the spacetime must take $\mathscr{I}^+$ to itself, the vector
fields that they determine on $\mathscr{I}^+$ must be tangent to
$\mathscr{I}^+$. In particular, the vector fields that the translations yield
on $\mathscr{I}^+$ are all proportional to the tangent of the \emph{null}
geodesic generators of $\mathscr{I}^+$. However, in contrast to the spatial
infinity case, the factor of proportionality is \emph{not} constant, but
rather is a linear combination of the first four ordinary spherical harmonics.

In fact, these vector fields, the so-called Bondi--Metzner--Sachs (BMS)
translational vector fields, can be characterized completely in terms of the
\emph{conformal structure} of $\mathscr{I}^+$. This yields the notion of
asymptotic translations \emph{even in general spacetimes admitting
$\mathscr{I}^+$} \cite{Sachs,Ge,NT80,PR2}; and these coincide with the
(equivalence classes of the) asymptotically constant solutions of the
\emph{asymptotic Killing equation} \cite{GeWi}. The BMS translational vector
fields can be characterized by their \emph{Weyl} or \emph{2-component spinor
constituents}, too: they are proportional to the constituent spinor of the
tangent of the null geodesic generators of $\mathscr{I}^+$, and the factor
of proportionality is a linear combination of spin-weight $1/2$
spherical harmonics. These can be recovered as the solutions of various
linear partial differential equations on the cut. (For a list of these, see
the appendix of \cite{Sz01}.)

\subsubsection{The superpotential}
\label{sub-1.1.3}

In the literature several different forms of the integrand in (\ref{eq:1.1})
are known both for the ADM \cite{ADM} and TBS
\cite{Trautman58,Bondietal,Sachs} energy-momenta: they can be given
(1) by certain expansion coefficients in the asymptotic expansion of the
components of the (spatial or spacetime) metric as a series of $1/r$ in an
asymptotic Cartesian or retarded null coordinate system; or
(2) by the traditional superpotential of some classical (e.g. Einstein's)
energy-momentum pseudotensor (or, in the tetrad formalism of the theory, by
an $SO(1,3)$ gauge dependent energy-momentum complex) in some appropriate
coordinate system (or Lorentz frame).
Also,
(3) one can write the spacetime metric as the sum of the flat spacetime
metric $\bar g_{ab}$ and some `correction field' $\gamma_{ab}$, and then
rewrite the exact Einstein equations by the linearized Einstein tensor, built
from $\gamma_{ab}$ on the background $\bar g_{ab}$, and an effective
energy-momentum tensor. The $U$ in (\ref{eq:1.1}) is the contraction of the
superpotential for this effective energy-momentum tensor and a translational
Killing field of the background metric.
(4) The 2-surface integral (\ref{eq:1.1}) can also be sought in the
form of the boundary term in the Hamiltonian of the theory, in which the
lapse and the shift are the timelike and spacelike part of $K^a$, respectively.
(5) These total energy-momenta can also be written as a Komar or linkage
integral, based on the vector field $K^a$ (see \cite{WiTa});
or,
(6) using the field equations, they can be re-expressed by certain parts of
the curvature tensor.

However, (7) there is a `universal superpotential', built from a Dirac spinor
or a pair of 2-component spinors, by means of which the classical (e.g. the
pseudotensorial) superpotentials could be recovered (see the Introduction);
and \emph{both} the ADM and TBS energy-momenta can be re-expressed. This is
the Nester--Witten 2-form \cite{Ne,Witt}, given in terms of any two
2-component spinors \cite{HT} by

\begin{equation}
u(\lambda,\bar\mu)_{ab}:=\frac{\rm i}{2}\bigl(\bar\mu_{A'}\nabla_{BB'}
\lambda_A-\bar\mu_{B'}\nabla_{AA'}\lambda_B\bigr). \label{eq:1.2}
\end{equation}
Although this is a complex-valued 2-form, its integral on any closed,
orientable 2-surface ${\cal S}$ defines a \emph{Hermitian} bilinear form on
the space of the spinor fields on ${\cal S}$.

To ensure the existence of the integrals in (\ref{eq:1.1}), and, in the
traditional formulation, also the independence of the ADM energy-momentum
of the coordinate system/background Minkowski metric, non-trivial fall-off
conditions should be imposed both on the matter fields and the geometric
data $(h_{ab},\chi_{ab})$ on the hypersurface. The latter should tend
asymptotically to the trivial flat or hyperboloidal data on the hypersurfaces
$\Sigma_t$ and $\Sigma_\tau$, respectively.

The key properties of the ADM and TBS energy-momenta are that they are
future pointing and timelike vectors (a property we will call `positivity'),
provided the energy-momentum tensor of the matter fields satisfies the
dominant energy condition on the spacelike hypersurface $\Sigma$ whose
boundary at infinity is ${\cal S}$; and the vanishing of these energy-momenta
is equivalent to the flatness of the domain of dependence $D(\Sigma)$ of
$\Sigma$ and the vanishing of the matter fields (`rigidity')
\cite{ScY,Witt,Ne,HT,RT}.  These results together, known as the positive
energy or positive mass theorem, hold true even in the presence of black holes
\cite{Gibbonsetal}. Probably the simplest proof of this theorem is based on
the use of spinors, the superpotential (\ref{eq:1.2}) and the Witten-type
gauge condition for the spinor fields on the hypersurface $\Sigma$ \cite{RT}. 
The rigorous mathematical proof of the existence and uniqueness of the 
solution of the Witten equation on \emph{asymptotically flat} spacelike 
hypersurfaces could be based on the techniques of \cite{ChCh}. The analogous 
results on \emph{asymptotically hyperboloidal} hypersurfaces are given in 
\cite{SzT}. 

The spacetimes describing the history of \emph{closed} universes are defined
to be those globally hyperbolic spacetimes that admit \emph{compact} Cauchy
surfaces. Clearly, these do \emph{not} represent the gravitational field of
localized sources, and total energy-momentum of the form (\ref{eq:1.1})
cannot be associated with them. Nevertheless, their \emph{total mass} can
be introduced using the (common) \emph{3-surface integral form} of the ADM
and TBS energy-momenta. Although its non-negativity is trivial by construction
(given the assumption that the dominant energy condition holds), still it has
a non-trivial rigidity property: this total mass is zero precisely when the
spacetime is locally flat, the matter fields are vanishing and the Cauchy
surface is a 3-torus \cite{Sz12,Sz13a}.

\subsection{On total energy-momentum with $\Lambda<0$}
\label{sub-1.2}

\subsubsection{The domain of integration}
\label{sub-1.2.1}

\begin{figure}[ht]
\begin{center}
\includegraphics[width=0.35\textwidth]{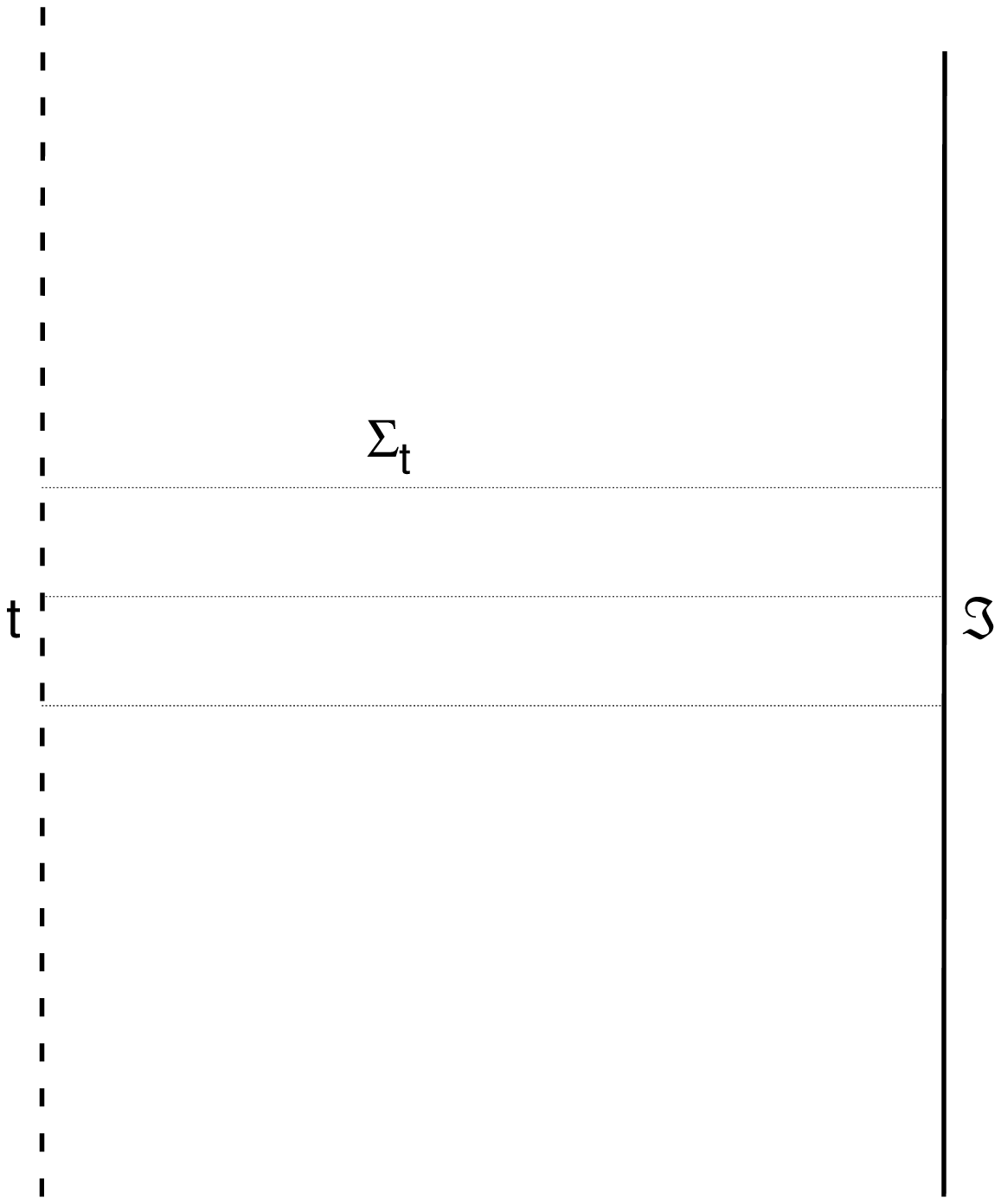}
\includegraphics[width=0.35\textwidth]{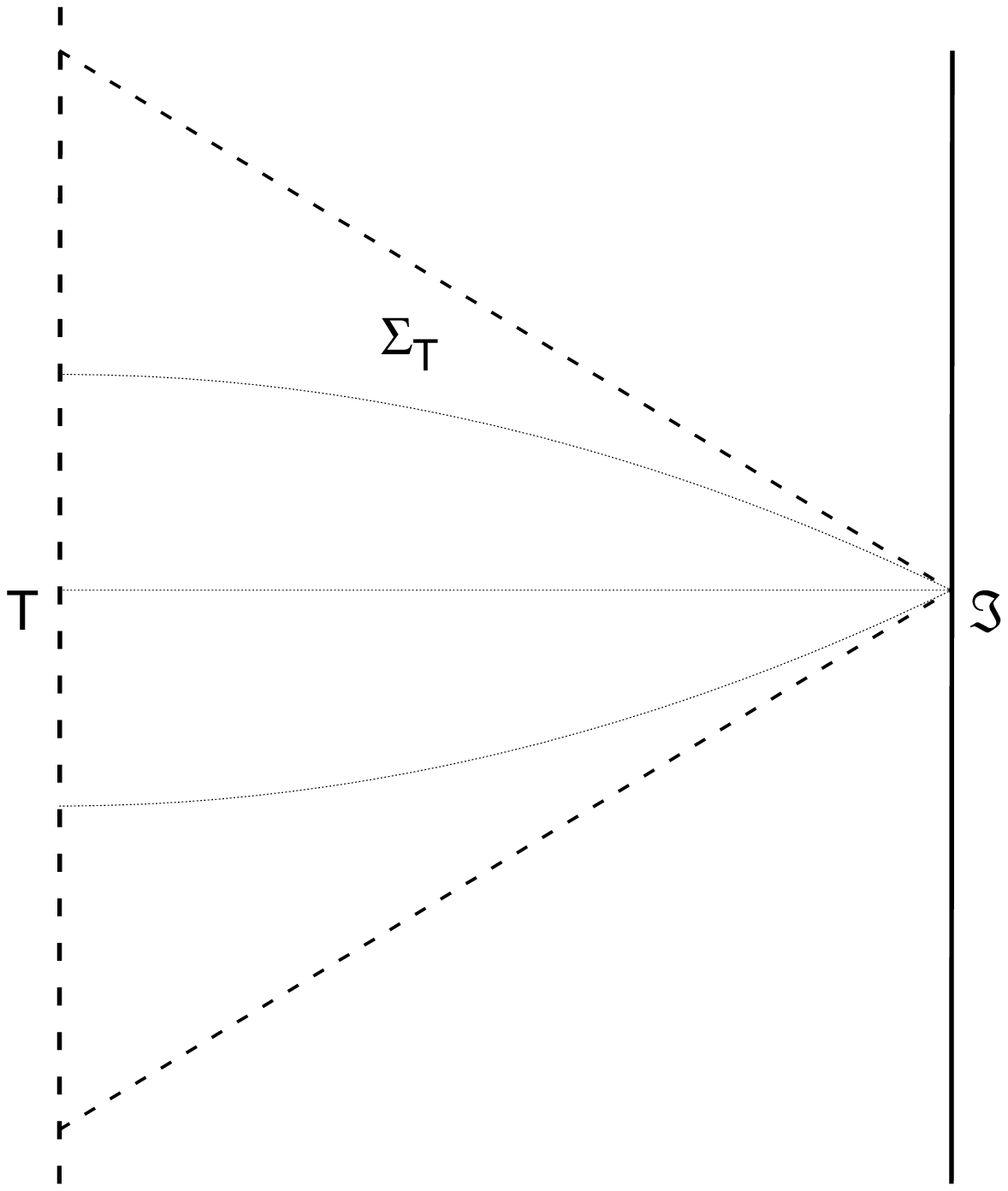}
\caption{\label{fig:AdeSitter}
Foliations of the anti-de Sitter spacetime: i. The hypersurfaces $\Sigma_t$ 
form a global foliation. The leaves $\Sigma_t$ are partial Cauchy surfaces 
with $\mathbb{R}^3$ topology, and these foliate the (timelike) conformal 
boundary $\mathscr{I}$, too. The $t={\rm const}$ hypersurfaces in the global 
coordinates of the anti-de Sitter spacetime are like this. These hypersurfaces 
are intrinsically hyperboloidal and extrinsically flat. 
ii. The hypersurfaces $\Sigma_T$ foliate neither the whole spacetime nor its 
conformal boundary $\mathscr{I}$. They foliate only a globally hyperbolic 
open subset. These hypersurfaces are intrinsically asymptotically 
hyperboloidal, and their extrinsic curvature is asymptotically proportional 
to the intrinsic metric. Such a $T$, whose level sets are the hypersurfaces 
$\Sigma_T$, is the time coordinate in the FRW form of the anti-de Sitter 
metric. (See also Figure 20 of \cite{HE}.)} 
\end{center}
\end{figure}

If $\Lambda<0$, then the gravitational field of a \emph{localized} source is 
expected to be represented by a spacetime of asymptotically constant negative 
curvature, i.e. by some `asymptotically anti-de Sitter' spacetime. The 
anti-de Sitter spacetime itself \cite{HE} is the universal covering space of 
the vacuum spacetime with constant negative curvature $R=4\Lambda<0$. Its 
conformal boundary $\mathscr{I}$ is a \emph{timelike} hypersurface in its 
conformal completion with topology $S^2\times\mathbb{R}$. In the standard 
\emph{global} coordinates $(t,r,\theta,\phi)$ \cite{HE} the $t={\rm const}$ 
spacelike hypersurfaces, denoted by $\Sigma_t$, are \emph{partial} Cauchy 
surfaces and intersect $\mathscr{I}$ in 2-spheres. They provide a foliation 
of $\mathscr{I}$ as well\footnote{The anti-de Sitter line element can be 
rewritten in FRW form with $k=-1$, but the leaves of this latter 
foliation would foliate only globally hyperbolic proper subsets of the 
spacetime, say $D(\Sigma_t)$ for given $t$ (see 
Fig.~\ref{fig:AdeSitter}.ii.).} (see Fig.~\ref{fig:AdeSitter}.i.). The 
induced metric on them is of constant negative curvature with scalar 
curvature ${\cal R}=2\Lambda$ and vanishing extrinsic curvature, i.e. these 
hypersurfaces are \emph{intrinsically hyperboloidal} and \emph{extrinsically 
flat}.

The traditional definition of asymptotically anti-de Sitter spacetimes is
based on the decomposition of the spacetime metric into the sum of the
anti-de Sitter metric $\bar g_{ab}$ and some `correction term' $\gamma_{ab}$
\cite{AD}; but the latter tensor field should be required to satisfy
appropriate fall-off conditions. However, adapting the key ideas of the
conformal approach to infinity to the present case, the asymptotically
anti-de Sitter spacetimes could be defined in a fully geometric way as in
\cite{AM}, by the conformal compactifiability of the spacetime. Since
$\Lambda<0$ the conformal boundary $\mathscr{I}$ is \emph{timelike}, the
trace-free part of its extrinsic curvature is vanishing and the trace can be
made to be vanishing in some conformal gauge. However, its intrinsic metric
at this point is not restricted further, and could still be \emph{completely
general}. Thus, to restrict the asymptotic properties of the spacetime to be
similar to that of the anti-de Sitter, the conformal boundary $\mathscr{I}$
is usually required to be topologically $S^2\times\mathbb{R}$ and
\emph{intrinsically conformally flat} \cite{AM}. The latter condition is
called the `reflective boundary condition' \cite{Ha83}, and is equivalent to 
the vanishing of the magnetic part of the conformally rescaled Weyl curvature 
\cite{Tod84}, determined by the spacelike normal of $\mathscr{I}$, on 
$\mathscr{I}$. Nevertheless, this boundary condition excludes any 
gravitational energy-flux through the conformal boundary, and, as Friedrich 
stresses (see \cite{Friedrich}, and especially \cite{Friedrich14a}), this 
boundary condition is one choice among many. Other `natural' boundary 
conditions, when incoming and/or outgoing energy flux through $\mathscr{I}$ 
is allowed, are also possible \cite{Friedrich14a}.

The domain ${\cal S}$ of integration in (\ref{eq:1.1}) is a closed spacelike
2-surface, which may also be called a `cut', in $\mathscr{I}$. In the exact
anti-de Sitter spacetime such a cut is, for example, the intersection
$\Sigma_t\cap\mathscr{I}$. Since, however, $\mathscr{I}$ is \emph{timelike},
there is no way to distinguish the ADM and TBS type energy-momenta: the
specific properties of the total energy-momentum associated with the cuts of
$\mathscr{I}$ depend crucially on the boundary conditions on $\mathscr{I}$.

\subsubsection{The asymptotic translations}
\label{sub-1.2.2}

The Lie algebra of the Killing fields is the anti-de Sitter Lie algebra
$so(2,3)$. Since, however, $so(2,3)$ is \emph{semi-simple}, there is no way
to single out `translations' in a canonical \emph{algebraic} way; nor, since
this spacetime does not admit any non-trivial constant vector field, is there
a way to single out `translations' in a canonical \emph{geometric} manner
either. Nevertheless, the coordinate $t$ is timelike and the components of
the metric in the global coordinates $(t,r,\theta,\phi)$ are independent of
$t$, so the Killing field $(\partial/\partial t)^a$ is often interpreted as
the \emph{time translational} Killing field (though, for example, it is
\emph{not} a geodesic vector field). The Killing fields of the spacetime
extend to \emph{conformal Killing fields} on infinity, and, in particular,
$(\partial/\partial t)^a$ extends to a timelike one on $\mathscr{I}$.

If, however, the `asymptotically anti-de Sitter spacetimes' are defined
simply by the existence of a (timelike) conformal boundary $\mathscr{I}$
(possibly with $S^2\times\mathbb{R}$ topology) but \emph{without} any further
condition on its intrinsic conformal geometry, then in general $\mathscr{I}$
does not admit any conformal isometry. Hence, the spacetime cannot admit any
`asymptotic Killing vector' either. On the other hand, the requirement of the
intrinsic conformal flatness of $\mathscr{I}$, i.e. the reflective boundary
condition, guarantees the maximal number of conformal Killing fields, and
their Lie algebra is isomorphic to the anti-de Sitter algebra $so(2,3)$
\cite{AM}.

\subsubsection{The superpotential}
\label{sub-1.2.3}

As in the $\Lambda=0$ case (in subsection \ref{sub-1.1.3}), several different
superpotentials can (and, in fact, have been) used in (\ref{eq:1.1}) to
associate energy-momentum with cuts of the timelike conformal boundary
$\mathscr{I}$. For example, that could be based on the use of the effective
energy-momentum tensor built from $\gamma_{ab}$ mentioned in (3) of subsection
\ref{sub-1.1.3} \cite{AD}; or on another explicitly given superpotential
\cite{ChruscielNagy}, obtained from a Hamiltonian. Also, the precise fall-off
conditions for $\gamma_{ab}$ are determined which make the resulting
expressions unambiguously defined. (Without this, the expression of \cite{AD}
would suffer from coordinate ambiguities.) The integrand used in \cite{AM} is
the electric part of the rescaled Weyl curvature, contracted with a conformal
Killing vector of $\mathscr{I}$; and this was shown in \cite{Ke} to coincide
with the `renormalized' quasi-local mass of Penrose \cite{PR2} (after
subtracting the cosmological constant term), calculated on a cut of
$\mathscr{I}$. The latter two are manifestly coordinate-free, and all these
expressions, when they are well-defined, give the same result. The
Nester--Witten 2-form (\ref{eq:1.2}) in its `renormalized' form is used in
\cite{Gibbonsetal}, where a Witten-type proof of the positivity of energy,
and also of rigidity, are also given. In all these investigations the
intrinsic conformal flatness of $\mathscr{I}$ was assumed.

\subsection{The need for total energy-momenta with $\Lambda>0$}
\label{sub-1.3}

The simplest explanation of the \emph{observed} accelerating expansion of
the Universe \cite{Riessetal,Perlmutteretal} is the presence of a strictly
positive cosmological constant in Einstein's field equations. Thus, the future
history of our Universe is asymptotically de Sitter. Also, the positivity
of $\Lambda$ is the basis of the conformal cyclic cosmological (CCC) model
of Penrose \cite{Pe10}. In the analysis of the asymptotic structure of these
spacetimes, the total energy-momenta could still provide useful tools. In
fact, the need (and three potential expressions) for the total TBS type
energy was already raised by Penrose in \cite{Pe11}. Since then several
papers devoted to the question of total energy-momentum in asymptotically de
Sitter spacetimes have appeared. The aim of the present paper is to give a
review of these results.

In section \ref{sec-2} we review the properties of general asymptotically
de Sitter spacetimes; and discuss the important special spacetimes that are
asymptotic to de Sitter, with special emphasis on their foliations,
symmetries and the fields on the background de Sitter spacetime. Then, in
section \ref{sec-3}, we discuss ADM type energy-momentum expressions, while
sections \ref{sec-4} and \ref{sec-5} are devoted to the TBS type expressions
and a suggestion for the total mass in closed universes, respectively. The
signature of the spacetime metric is chosen to be $(+,-,-,-)$, and Einstein's
equations are written in the form $R_{ab}-\frac{1}{2}Rg_{ab}=-\kappa T_{ab}-
\Lambda g_{ab}$ with $\kappa=8\pi G$.


\section{The structure of asymptotically de Sitter spacetimes}
\label{sec-2}

\subsection{The de Sitter spacetime}
\label{sub-2.1}

\subsubsection{Foliations}
\label{sub-2.1.1}

\begin{figure}[ht]
\begin{center}
\includegraphics[width=0.35\textwidth]{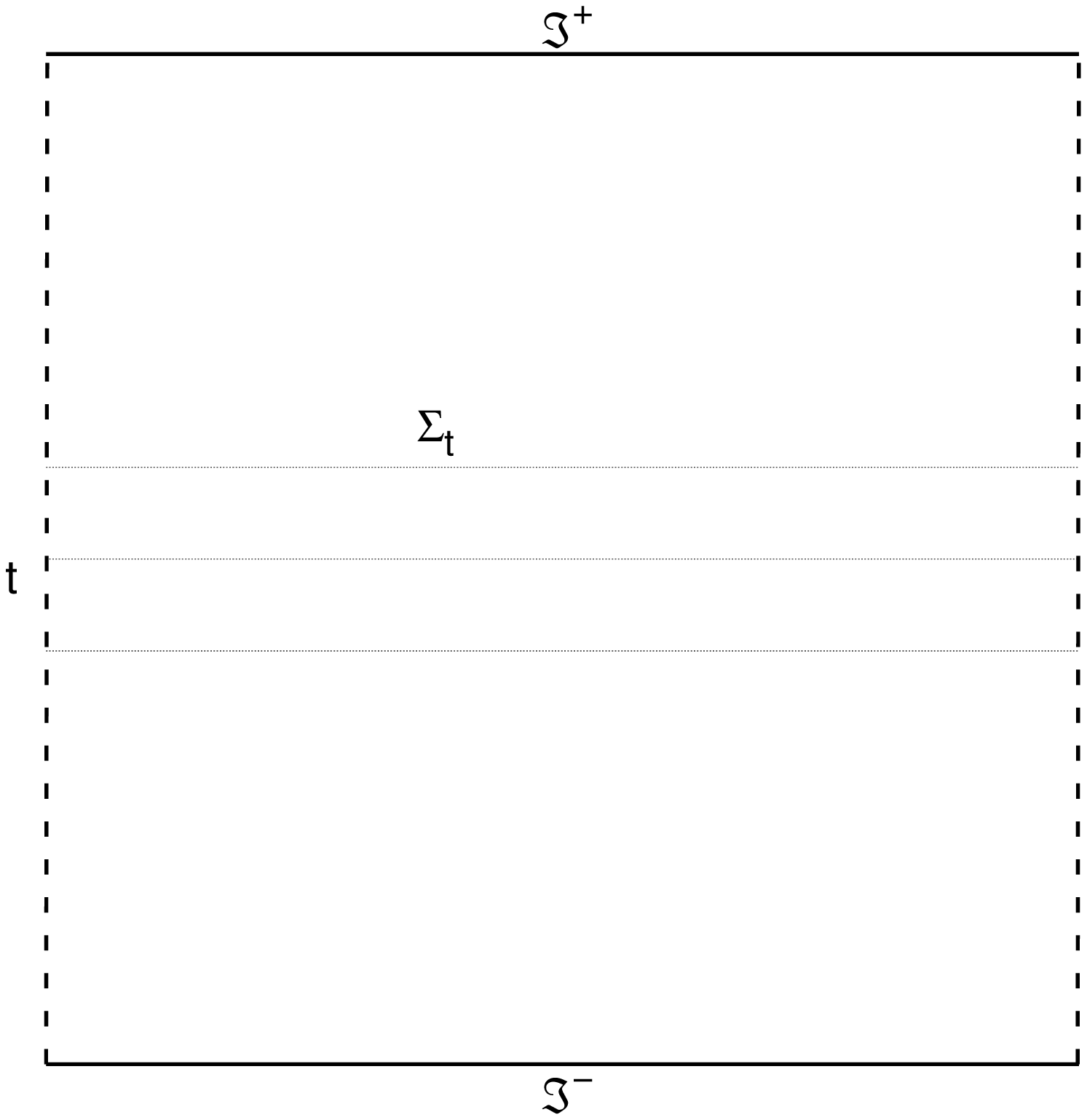}
\includegraphics[width=0.35\textwidth]{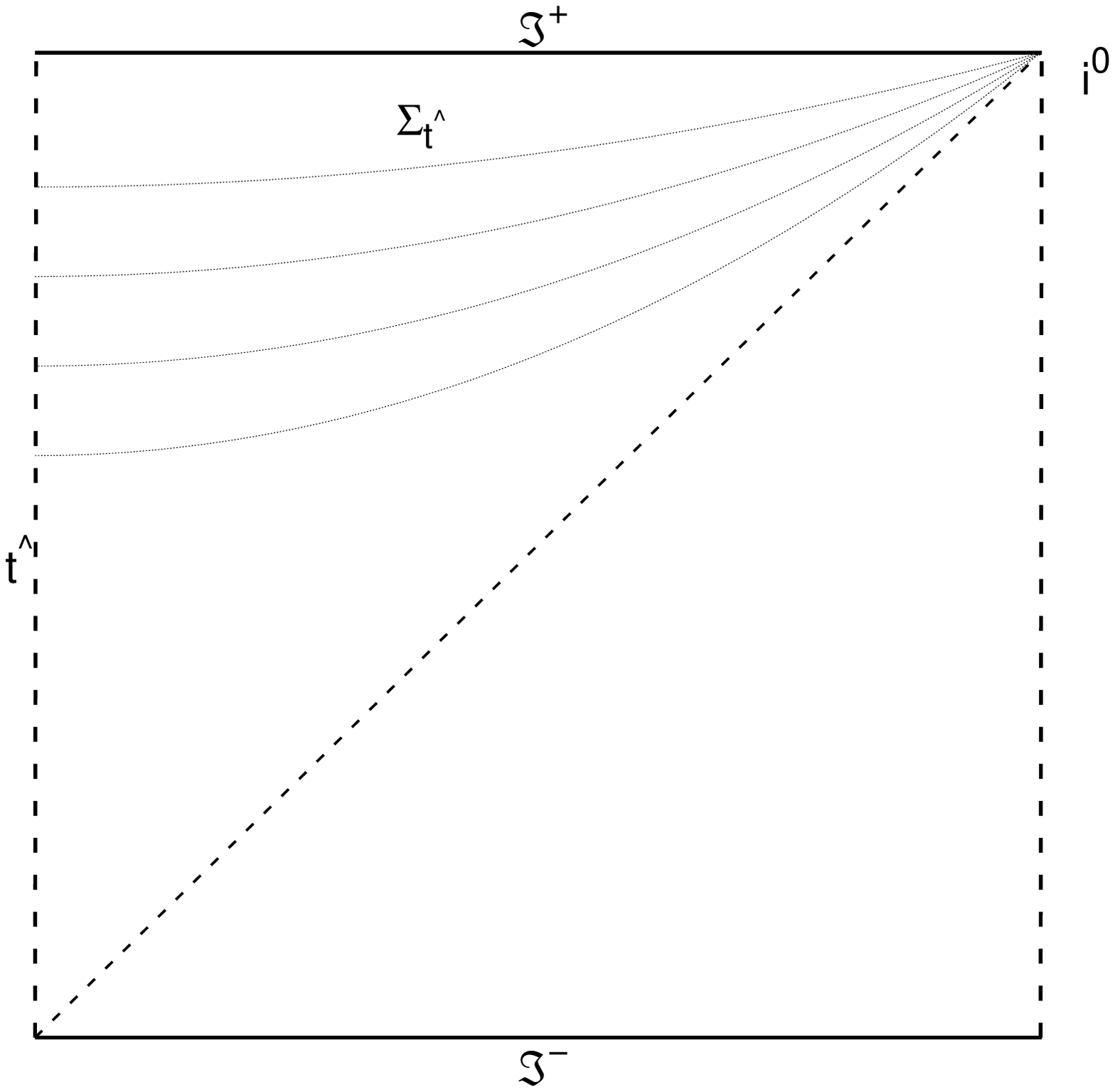}
\caption{\label{fig:deSitter-a}
Foliations of the de Sitter spacetime: i. The hypersurfaces $\Sigma_t$ 
form a global foliation. The leaves $\Sigma_t$ are global Cauchy surfaces 
with $S^3$ topology. The $t={\rm const}$ hypersurfaces in the global 
coordinates of the de Sitter spacetime are like this, which are metric 
spheres. 
ii. The hypersurfaces $\Sigma_{\hat t}$ foliate the `steady state' part of (or 
`Poincar\'e patch' in) the (half) de Sitter spacetime. Their topology is 
$\mathbb{R}^3$, they are intrinsically asymptotically flat and their 
extrinsic curvature is asymptotic to a nonzero value. Their one-point 
compactification yields a `spatial infinity' $i^0$, which is a point of the 
future conformal boundary $\mathscr{I}^+$. These hypersurfaces do \emph{not} 
foliate $\mathscr{I}^+$. (See also Figure 17 of \cite{HE}.)} 
\end{center}
\end{figure}

The de Sitter spacetime is the constant positive curvature solution of the 
vacuum Einstein equations with scalar curvature $R=4\Lambda>0$. (For a 
summary of its key geometric properties, see \cite{Tod15}, and for a detailed 
discussion of its global structure, especially its Penrose diagram, see 
\cite{HE}.) Its line element in the \emph{global} coordinates $(t,r,\theta,
\phi)$ is 

\begin{equation*}
ds^2=dt^2-\alpha^2\cosh^2(t/\alpha)\Bigl(dr^2+\sin^2r\bigl(\, d\theta^2+\sin^2
\theta \, d\phi^2\bigr)\Bigr),
\end{equation*}
where $\alpha^2:=3/\Lambda$, and the range of the coordinates are $t\in
\mathbb{R}$, $r\in[0,\pi]$ and $(\theta,\phi)\in S^2$. This metric has the 
FRW form with $k=1$, scale function $S(t)=\alpha\cosh(t/\alpha)$ and 
foliation by \emph{global} Cauchy surfaces $\Sigma_t:=\{\,t={\rm const}\,\}$ 
with $S^3$ topology (see Fig.~\ref{fig:deSitter-a}.i.). Then, in the  
coordinates $(\tau,r,\theta,\phi)$ with $\tau:=2\arctan(\exp(t/\alpha))-
\pi/2$, the future/past conformal boundary, $\mathscr{I}^\pm$, is given by 
$\tau=\pm\pi/2$ and 

\begin{equation*}
ds^2=S^2\Bigl(d\tau^2-dr^2-\sin^2r\bigl(d\theta^2+\sin^2\theta\,d\phi^2\bigr)
\Bigr).
\end{equation*}

The (half) de Sitter spacetime, often called the `steady state universe' or 
`Poincar\'e patch', has another foliation with \emph{intrinsically flat} 
spacelike hypersurfaces as well \cite{HE} (see Fig.~\ref{fig:deSitter-a}.ii.). 
These are the $\Sigma_{\hat t}:=\{\,\hat t={\rm const}\,\}$ hypersurfaces, 
where 

\begin{eqnarray}
&{}&\hat t:=\alpha\ln\Bigl(\sinh(\frac{t}{\alpha})+\cosh(\frac{t}{\alpha})
 \cos r\Bigr), \hskip 20pt
 \hat x:=\frac{\alpha}{\tanh(\frac{t}{\alpha})+\cos r}\sin r\,\sin
 \theta\,\cos\phi, \nonumber \\
&{}&\hat y:=\frac{\alpha}{\tanh(\frac{t}{\alpha})+\cos r}\sin r\,\sin
 \theta\,\sin\phi, \hskip 20pt
 \hat z:=\frac{\alpha}{\tanh(\frac{t}{\alpha})+\cos r}\sin r\,\cos
 \theta. \label{eq:2.1}
\end{eqnarray}
This coordinate system covers only the `steady state' part of the de Sitter 
spacetime, for which $\sinh(t/\alpha)+\cosh(t/\alpha)\cos r>0$, and the line 
element takes the form 

\begin{equation*}
ds^2=d\hat t^2-\exp(2\hat t/\alpha)\Bigl(d\hat x^2+d\hat y^2+d\hat z^2\Bigr).
\end{equation*}
Thus, this line element has the FRW form with $k=0$ and scale function 
$S(\hat t)=\exp(\hat t/\alpha)$. Hence, the extrinsic curvature of the 
hypersurfaces $\Sigma_{\hat t}$ is a pure trace, and the mean curvature is 
the \emph{positive constant} $\chi=3/\alpha$. All these hypersurfaces reach 
$\mathscr{I}^+$ at the point $r=\pi$, anti-podal to the origin, where 
they are, in fact, \emph{tangential} to $\mathscr{I}^+$. Thus, the anti-podal 
point provides a one-point-compactification of these hypersurfaces, and this 
can be interpreted as the `spatial infinity' of the flat 3-spaces. Clearly, 
this foliation is analogous to the foliation of the Minkowski spacetime by 
the $t={\rm const}$ hyperplanes. Nevertheless, the hypersurfaces $\Sigma
_{\hat t}$ are \emph{extrinsically hyperboloidal}; and none of these 
hypersurfaces is a global Cauchy surface, because no future inextendible 
non-spacelike curve terminating at the anti-podal point intersects any of 
these $\Sigma_{\hat t}$. These are Cauchy surfaces only for the steady state 
part of the de Sitter spacetime. 

\begin{figure}[ht]
\begin{center}
\includegraphics[width=0.35\textwidth]{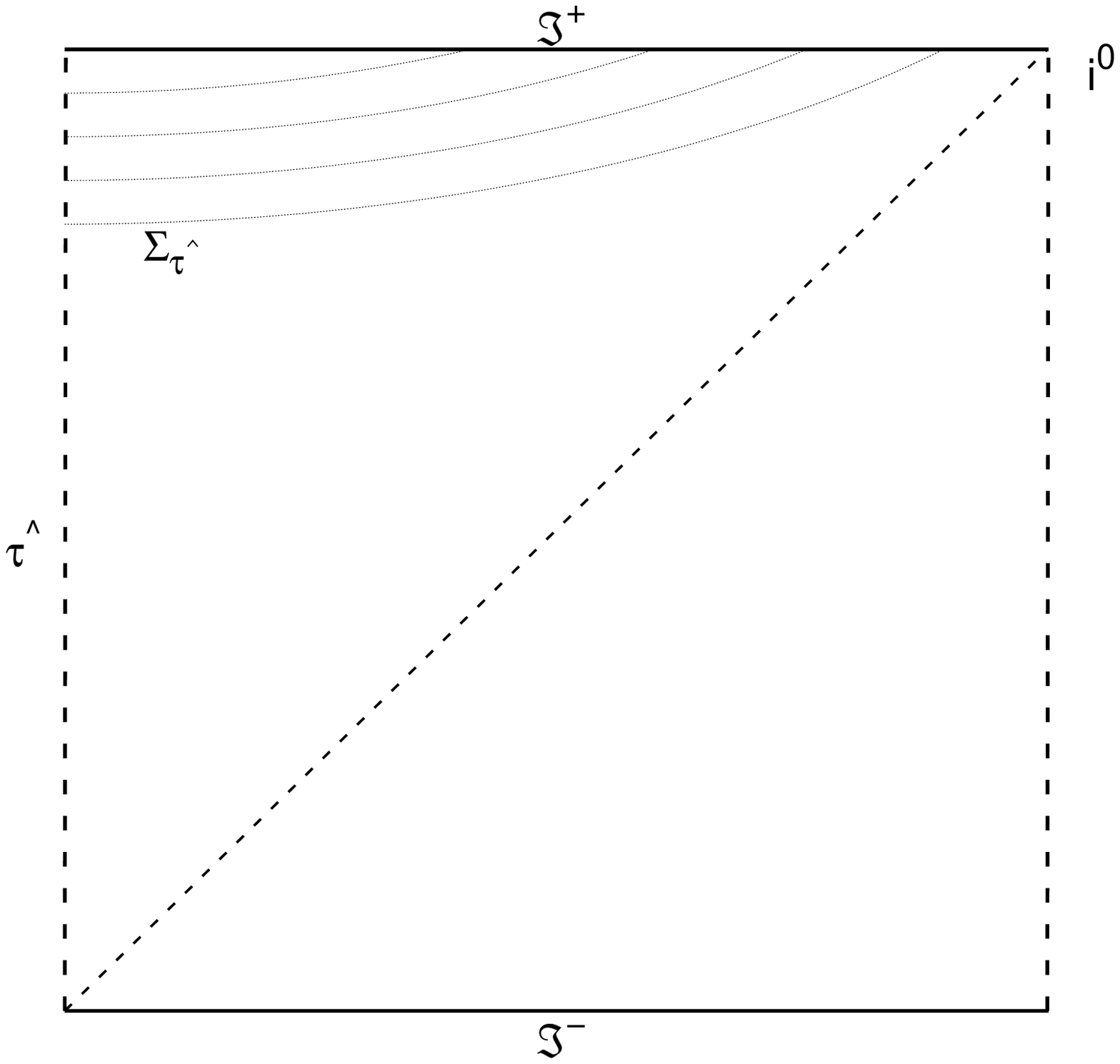}
\includegraphics[width=0.35\textwidth]{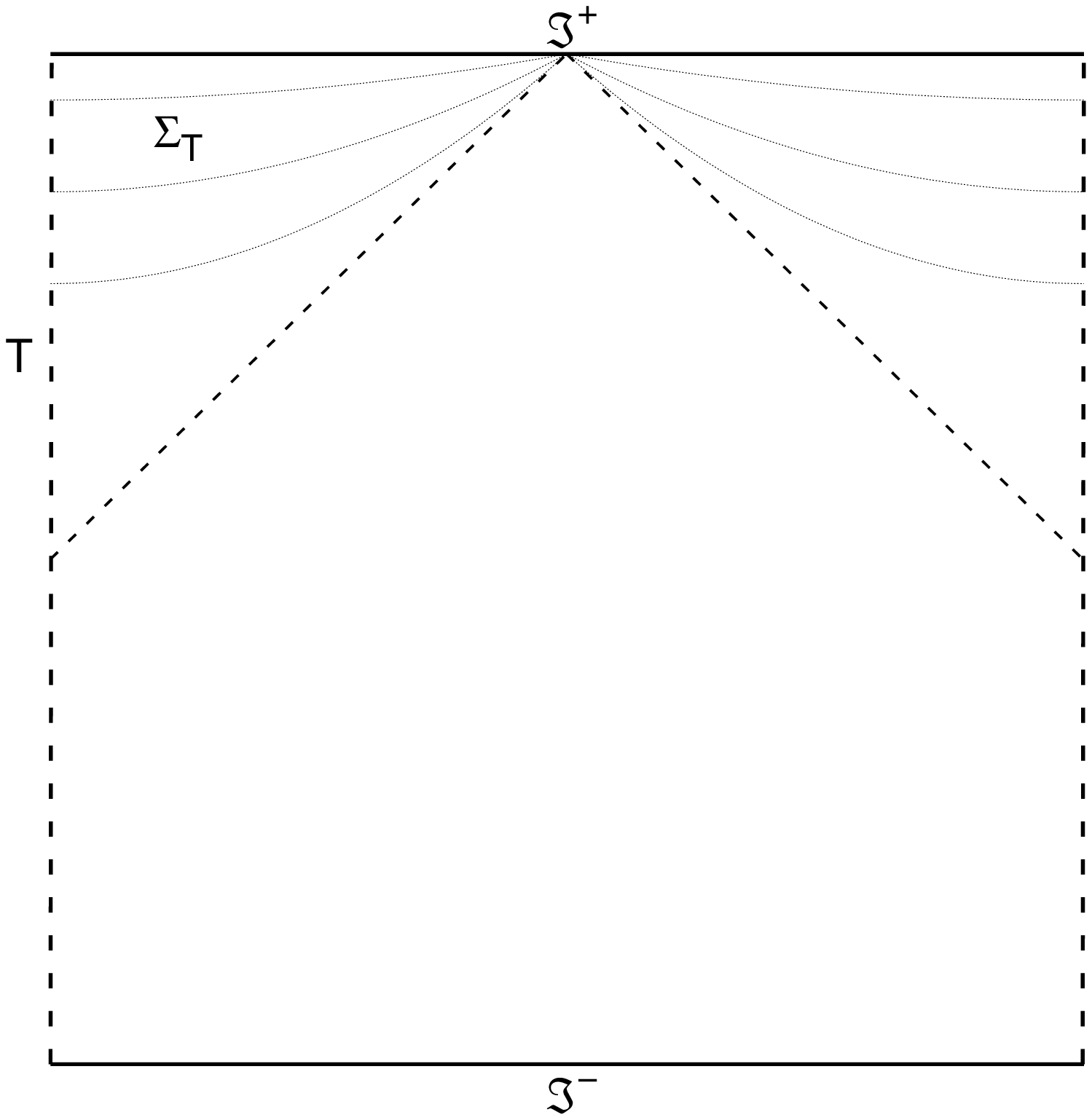}
\caption{\label{fig:deSitter-b}
Foliations of the de Sitter spacetime: iii. The hypersurfaces $\Sigma
_{\hat\tau}$ foliate \emph{both} the `steady state' part of the de Sitter 
spacetime and the (spacelike) future conformal boundary $\mathscr{I}^+$ 
(minus $i^0$). The leaves $\Sigma_{\hat\tau}$ are intrinsically asymptotically 
hyperboloidal, and their extrinsic curvature is asymptotically proportional 
to the intrinsic metric. 
iv. The hypersurfaces $\Sigma_T$ foliate the disjoint union of the past 
domain of dependence of the $r\in[0,\pi/2)$ and the $r\in(\pi/2,\pi]$ 
hemispheres of the conformal boundary $\mathscr{I}^+\approx S^3$. They do 
\emph{not} provide a foliation of the conformal boundary itself. } 
\end{center}
\end{figure}

In contrast to this foliation, a neighbourhood of $\mathscr{I}^+-
\{(\pi,0,0)\}$ can be foliated by spacelike hypersurfaces which foliate 
$\mathscr{I}^+-\{(\pi,0,0)\}$, too (see e.g. \cite{SzT}). Such a foliation 
can be based on the coordinate system $(u,w,\theta,\phi)$, where, in terms 
of the global coordinates above, $u:=\alpha(\tau-r)$ and $w:=(\pi/2-\tau)/
\alpha$. Then $\mathscr{I}^+$ is just the $w=0$ hypersurface, and the $u=0$ 
`origin cut' of $\mathscr{I}^+$ is the $r=\pi/2$ maximal 2-sphere. Then, for 
a fixed positive $W$ and $\hat\tau:=u-Ww$, let us define $\Sigma_{\hat\tau}:=
\{\hat\tau={\rm const}\}$. These are spacelike hypersurfaces, $-\alpha\pi/2
\leq\hat\tau\leq\alpha\pi/2$, and $\Sigma_{\hat\tau}\cap\mathscr{I}^+$ is the 
cut $r=\pi/2-\hat\tau/\alpha$. (Although these hypersurfaces are not smooth 
at $r=0$ [in the original coordinates], these conical singularities can be 
smoothed out.) Hence, $\{\Sigma_{\hat\tau}\}$ provides, in fact, 
a foliation of the conformal boundary\footnote{The hypersurfaces $\Sigma
_T:=\{T={\rm const}\}$, given in the coordinates $(T,R,\theta,\phi)$ defined 
by $\cosh(T/\alpha):=\cosh(t/\alpha)\cos r$, $\sinh(T/\alpha)\cosh R:=\sinh
(t/\alpha)$ (see e.g. \cite{Luoetal}), foliate only the disjoint union of the 
past domain of dependence of the $r\in[0,\pi/2)$ and the $r\in(\pi/2,\pi]$ 
pieces of $\mathscr{I}^+$. Moreover, all of these hypersurfaces intersect 
$\mathscr{I}^+$ in the 2-sphere $r=\pi/2$, rather than giving a foliation of 
$\mathscr{I}^+$ (see Fig.~\ref{fig:deSitter-b}.iv.).} (see 
Fig.~\ref{fig:deSitter-b}.iii.). One can show \cite{SzT} that in the de 
Sitter spacetime their intrinsic geometry is asymptotically hyperboloidal 
and their extrinsic curvature is asymptotically proportional to their 
intrinsic metric. 

The analog of the three special foliations shown by Figures 
\ref{fig:deSitter-a}.i., \ref{fig:deSitter-a}.ii. and 
\ref{fig:deSitter-b}.iii. are used in asymptotically de Sitter spacetimes in 
a total mass construction in closed universes (section \ref{sec-5}), and in 
the definition of ADM type (section \ref{sec-3}) and of TBS type 
energy-momenta (section \ref{sec-4}), respectively. The foliation shown by 
Fig.~\ref{fig:deSitter-b}.iv. (like the ones shown by 
Fig.~\ref{fig:Minkowski}.ii. in the Minkowski and by 
Fig.~\ref{fig:AdeSitter}.ii. in the anti-de Sitter cases) does not seem to 
provide an appropriate framework in which the mass-loss property of a TBS 
type energy-momentum could potentially be proven.

\subsubsection{The Killing and conformal Killing fields}
\label{sub-2.1.2}

The de Sitter spacetime is of maximal symmetry: the Lie algebra of its
Killing and conformal Killing vectors is $so(1,4)$, the so-called de Sitter
Lie algebra, and $so(2,4)$, respectively. Since both are \emph{semi-simple},
generators of `translations' cannot be defined in a natural geometric or
algebraic way. On the other hand, there is an analytical characterization
of a basis of \emph{proper} conformal Killing vectors: these are
\emph{gradients} (see e.g. \cite{Tod15}) and any conformal Killing vector is
a sum of a gradient conformal Killing vector and a Killing vector.

Clearly, the six independent Killing vectors that are tangent to the
hypersurfaces $\Sigma_t$ are spacelike, and the remaining four are timelike
on some open subset, and spacelike on the interior of its complement.
However, since $\mathscr{I}^+$ is \emph{spacelike} and the isometries map
$\mathscr{I}^+$ to itself, \emph{every} spacetime Killing field must be
spacelike or zero on $\mathscr{I}^+$. For example, the Killing field

\begin{equation*}
K^{04}_e=-\cos r(\nabla_et)+\alpha\sinh\bigl(\frac{t}{\alpha}\bigr)\cosh
\bigl(\frac{t}{\alpha}\bigr)\sin r(\nabla_er),
\end{equation*}
given in the \emph{global} coordinates, is timelike precisely on the domain
where $\cos^2r>\tanh^2(t/\alpha)$ holds. In the $t\rightarrow\infty$ limit
the closure of this domain reduces to the two points $(0,0,0)$ and $(\pi,0,0)$
of the future conformal boundary, where $K^{04}_e$ vanishes. The remaining
three Killing vectors behave in a similar way. Hence, \emph{no} Killing
field is timelike on any open neighbourhood of some asymptotic end of the
hypersurfaces $\Sigma_{\hat t}$ of the intrinsically flat foliation of the
steady state part of the de Sitter spacetime. In particular, the Killing
field

\begin{equation}
K^a:=(\frac{\partial}{\partial\hat t})^a+\hat x(\frac{\partial}{\partial
\hat x})^a+\hat y(\frac{\partial}{\partial\hat y})^a+\hat z(\frac{\partial}
{\partial\hat z})^a, \label{eq:2.3}
\end{equation}
given in the coordinates (\ref{eq:2.1}) on the steady state part of the de
Sitter spacetime, is timelike precisely for $\exp(2\hat t/\alpha)(\hat x^2+
\hat y^2+\hat z^2)<1$. A simple (although less explicit) demonstration of the
non-existence of a \emph{globally timelike} Killing field in a neighbourhood
of $\mathscr{I}^+$ is given in \cite{Tod15}.

Clearly, one of the five \emph{conformal} Killing fields, namely $K^0_e=
\cosh(t/\alpha)(\nabla_et)$, is everywhere timelike, but the remaining four
are timelike only on some open subset. For example,

\begin{equation*}
K^4_e=\sinh\bigl(\frac{t}{\alpha}\bigr)\cos r(\nabla_et)-\alpha\cosh\bigl(
\frac{t}{\alpha}\bigr)\sin r(\nabla_er)
\end{equation*}
is timelike precisely when $\cosh^2(t/\alpha)\cos^2r>1$. In the $t\rightarrow
\infty$ limit this reduces to the disjoint union of the $r\in[0,\pi/2)$ and
$r\in(\pi/2,\pi]$ hemispheres of $\mathscr{I}^+$. This vector field is future
pointing on the $r\in[0,\pi/2)$, but past pointing on the $r\in(\pi/2,\pi]$
hemisphere. The remaining three conformal Killing fields behave in a similar
way. Rewriting the contravariant form of them in the $(\tau,r,\theta,\phi)$
coordinates we find that all these are orthogonal to $\mathscr{I}^+$ in the
conformal spacetime:

\begin{eqnarray}
&{}&K^a_0=\frac{1}{\alpha}(\frac{\partial}{\partial\tau})^a, \hskip 10pt
K^a_4\approx\cos r\, K^a_0, \label{eq:2.4} \\
&{}&K^a_1\approx\sin r\sin\theta\cos\phi\, K^a_0, \hskip 10pt
K^a_2\approx\sin r\sin\theta\sin\phi\, K^a_0, \hskip 10pt
K^a_3\approx\sin r\cos\theta\, K^a_0; \nonumber
\end{eqnarray}
where $\approx$ means `equal at the points of $\mathscr{I}^+$'. Therefore,
the structure of the conformal Killing vectors at the conformal boundary is
similar to that of the BMS translations at future null infinity in
asymptotically flat spacetimes, though now the number of the conformal
Killing vectors is five. Another (twistor theoretical) demonstration of this
statement is given in \cite{Tod15}, where all the 15 conformal Killing
vectors are constructed from the independent solutions of the twistor
equation, too. Using Dirac spinors, some of these results are also discussed
in \cite{KaTr}. Some of the ADM type energy-momenta are based on the use of
the asymptotic Killing vectors, while others depend on the use of the
asymptotic conformal Killing vectors.

\subsection{Asymptotically de Sitter spacetimes}
\label{sub-2.2}

\subsubsection{Fields on de Sitter background}
\label{sub-2.2.1}

In Yang--Mills theory on the de Sitter background the meaning of the vanishing
of the magnetic field strength on $\mathscr{I}^+$, i.e. the meaning of the
analog of the condition of the intrinsic conformal flatness of $\mathscr{I}
^+$, is investigated in \cite{ABK1}. It is shown that this condition removes
half the degrees of freedom of the Yang--Mills field; and all the ten de
Sitter fluxes, built from the energy-momentum tensor and the ten conformal
Killing fields of $\mathscr{I}^+$, vanish. A systematic investigation of the
Maxwell fields on and of the linear gravitational perturbation of the (steady
state part of the) de Sitter spacetime are given in \cite{ABK2}. The
covariant phase space method is used to obtain a Hamiltonian for them, in
which the `time translation' is a Killing field which is future pointing
timelike only in the intersection of the chronological past of the point
$(0,0,0)$ of $\mathscr{I}^+$ and the steady state universe, but spacelike on
the interior of the rest of the steady state universe. The resulting energy
flux through $\mathscr{I}^+$ can be \emph{negative and arbitrarily large}.
The requirement of the vanishing of the magnetic field strength, and of the
magnetic part of the rescaled Weyl curvature on $\mathscr{I}^+$, removes half
the degrees of freedom of the Maxwell and linear gravitational perturbations,
respectively, and yields vanishing energy flux through $\mathscr{I}^+$
\cite{ABK2}. Here the `energy' is defined by using the Killing field
(\ref{eq:2.3}). For a single, isolated source located at the origin, and in
the absence of incoming gravitational waves through the null boundary of the
steady state part of the de Sitter spacetime, which is the physical situation
of principal interest in the body of work here referred to, this energy is
proven to be positive. Under the same assumptions, and following a \emph{tour
de force} of calculation, in \cite{ABK3} a generalisation of the Einstein
quadrupole formula is derived and discussed. The summary of these results is
given in \cite{ABK4,Ash17}. For the calculation of the Penrose mass at a cut
of $\mathscr{I}^+$ in the linearized theory, see \cite{Tod15} and subsection
\ref{sub-4.1} below.

\subsubsection{The general definition of asymptotically de Sitter spacetimes}
\label{sub-2.2.2}

The geometric definition of asymptotically de Sitter spacetimes is based on
the idea of conformal compactifiability of the physical spacetime by
attaching a boundary hypersurface to it \cite{PR2}. Explicitly, it is assumed
that there is a manifold $\tilde M$ with non-empty boundary $\partial M$, a
Lorentzian metric $\tilde g_{ab}$ on $\tilde M$ and a smooth function $\Omega:
\tilde M\rightarrow[0,\infty)$ such that (1) $\tilde M-\partial M$ is
diffeomorphic to (and hence identified with) $M$; (2) $\tilde g_{ab}=\Omega^2
g_{ab}$ on $M$; (3) the boundary is just $\partial M=\{\Omega=0\}$ and
$\tilde\nabla_a\Omega$ is nowhere vanishing on $\partial M$; (4) $g_{ab}$
solves Einstein's equations, $R_{ab}-\frac{1}{2}Rg_{ab}=-\kappa T_{ab}-\Lambda
g_{ab}$, on $M$, where $\Lambda>0$; (5) $\tilde T^a{}_b:=\Omega^{-3}T^a{}_b$
can be extended to $\tilde M$ as a smooth field.

A systematic investigation of the asymptotic structure of spacetimes
satisfying these conditions is given in \cite{SzT,ABK1}. In particular, the
asymptotic form of spacetime metric, the Newman--Penrose spin coefficients,
the various pieces of the curvature and the energy-momentum tensor, the
geometry of asymptotically hyperboloidal hypersurfaces, etc; and also the
asymptotic field equations for them are determined. The conformal boundary
$\partial M$ is necessarily \emph{spacelike}, so it consists of the future
and past boundary, $\mathscr{I}^+$ and $\mathscr{I}^-$, respectively. (In
what follows, we concentrate only on $\mathscr{I}^+$. In a cosmological
setting $\mathscr{I}^-$ will usually not be present, being instead replaced
by an initial singularity.) The trace-free part of its extrinsic curvature
is vanishing and, in an appropriate conformal gauge, its trace can be made
zero; but \emph{its intrinsic conformal geometry is still completely
unrestricted}. Therefore, such a general asymptotically de Sitter spacetime
cannot admit any of the usual ten de Sitter Killing fields even
asymptotically, otherwise they would extend to conformal Killing fields on
$\mathscr{I}^+$. Thus, as the term is usually used, \emph{asymptotically de
Sitter spacetimes are not the spacetimes that are asymptotic to de Sitter
spacetime}. The latter form a very special subset of the former.

In fact, $\mathscr{I}^+$ admits ten linearly independent conformal Killing
vectors precisely when its intrinsic geometry is \emph{conformally flat};
which is equivalent to the vanishing of the magnetic part of the rescaled
Weyl tensor on $\mathscr{I}^+$. This condition
appears to be unreasonably strong: investigations of linear gravitational
perturbations of the de Sitter background reveal that this condition would
remove half the gravitational degrees of freedom \cite{ABK1,ABK2} (see
subsection \ref{sub-2.2.1} above) and in the full (nonlinear) theory this
is confirmed. The existence of a large number of spacetimes with positive
$\Lambda$ admitting a general future conformal boundary has been demonstrated
by Friedrich \cite{Friedrich14}. He shows that the free data for the Einstein
equations consist of a pair of symmetric tensors $(h_{ab},{\cal E}_{ab})$ with
$h_{ab}$ the Riemannian metric of $\mathscr{I}^+$ and ${\cal E}_{ab}$ the
electric part of the rescaled Weyl tensor at $\mathscr{I}^+$, subject to

\begin{equation*}
h^{ab}{\cal E}_{ab}=0, \hskip 20pt D^a{\cal E}_{ab}=0
\end{equation*}
and with the freedom $(h_{ab},{\cal E}_{ab})\mapsto(\Theta^2h_{ab},\Theta^{-1}
{\cal E}_{ab})$ for positive functions $\Theta$ on $\mathscr{I}^+$. (Here
$D_a$ is the covariant derivative on $\mathscr{I}^+$ defined by $h_{ab}$.)
These results, together with those in \cite{Friedrich,Friedrich14a}, show
the essential difference between the role of the boundary conditions in the
$\Lambda<0$ and $\Lambda>0$ cases: while in the $\Lambda<0$ case \emph{some}
boundary condition \emph{must} be specified on $\mathscr{I}$ to have a well
posed initial-boundary value problem for the field equations, in the
$\Lambda>0$ case similar boundary conditions on $\mathscr{I}^+$
\emph{restrict} the freely specifiable part of the Cauchy data.

A detailed discussion of the various global and asymptotic properties of
some special asymptotically de Sitter solutions (viz. the Schwarzschild--de
Sitter, Kerr--de Sitter, Vaidya--de Sitter and Friedman--Robertson--Walker)
is also given in \cite{ABK1}. In particular, the existence and properties of
the horizons, the global topology of the conformal boundary, the group of
the globally defined isometries and so on are clarified. Motivated by these
special cases, one may wish to impose further conditions, for example the
geodesic completeness of the conformal boundary.


\section{ADM type total energy-momenta}
\label{sec-3}

The definition of the ADM type energy-momenta are based on spacelike
hypersurfaces that extend to `spatial infinity'; i.e. which are analogous to
the foliation of the steady state part of the de Sitter spacetime by
intrinsically flat hypersurfaces, given by (\ref{eq:2.1}). Thus, strictly
speaking, such an ADM type energy-momentum is associated with a \emph{point}
$p$ of the conformal boundary $\mathscr{I}^+$ (still conveniently thought of
as `spatial infinity'), and the construction itself is based on a spacelike
hypersurface $\Sigma$ that is a Cauchy surface for the past domain of
dependence of $\mathscr{I}^+-\{p\}$ in the spacetime. These hypersurfaces are
\emph{tangent} to $\mathscr{I}^+$ at $p$, and hence $\Sigma$ is asymptotically
intrinsically flat and extrinsically asymptotically hyperboloidal. If the
intrinsic conformal geometry of $\mathscr{I}^+$ is homogeneous, e.g. when it
is conformally flat, then the energy-momentum is independent of $p$, but in
general it may depend on it. It is obvious how to generalize these ideas to
allow finitely many `spatial infinities'.

\subsection{The Abbott--Deser energy-momentum}
\label{sub-3.1}

The Abbott--Deser energy-momentum \cite{AD} is based on the decomposition
$g_{ab}=\bar g_{ab}+\gamma_{ab}$ of the spacetime metric into the sum of the
de Sitter metric $\bar g_{ab}$ and some tensor field $\gamma_{ab}$. Then
Einstein's equations can be written in the form ${}_LG_{ab}=\Lambda\gamma_{ab}
-\kappa t_{ab}$, where ${}_LG_{ab}$ denotes the linearized Einstein tensor on
the de Sitter background, built from $\gamma_{ab}$, and $t_{ab}$ is the sum of
the matter energy-momentum tensor and the correction to the linearized
Einstein tensor, being quadratic and higher order in $\gamma_{ab}$. By the
contracted Bianchi identity $\bar\nabla_at^a{}_b=0$ holds, where $\bar\nabla
_a$ is the covariant derivative in the de Sitter geometry and index raising
and lowering are defined by the background metric. Thus, $t_{ab}$ plays the
role of an \emph{effective energy-momentum tensor}. Therefore, for any
Killing vector $\bar K^a$ of the de Sitter background the contraction $\bar
K_at^{ab}$ is $\bar\nabla_b$-divergence-free, and hence the integral

\begin{equation}
{\tt Q}[\bar K]:=\int_\Sigma \bar K_at^{ab}\bar t_b{\rm d}\bar\Sigma
\label{eq:3.1}
\end{equation}
is independent of the hypersurface \emph{provided} it is finite and free of 
the ambiguity in the decomposition $g_{ab}=\bar g_{ab}+\gamma_{ab}$, which seems 
unlikely in general. Here $\bar t_a$ is the $\bar g_{ab}$-unit normal to, and 
${\rm d}\bar\Sigma$ is the induced volume element on $\Sigma$. In addition 
to the implicit assumption that ${\tt Q}[\bar K]$ is well defined, it is 
also assumed that the conformal boundary $\mathscr{I}^+$ is intrinsically 
conformally flat; otherwise the Killing fields $\bar K^a$ could not extend to 
conformal Killing vectors of $\mathscr{I}^+$ and they, as solutions of the 
asymptotic Killing equations in a neighbourhood of $\mathscr{I}^+$, would be 
ambiguously defined. By the conservation $\bar\nabla_b(\bar K_at^{ab})=0$ the 
integral ${\tt Q}[\bar K]$ can be rewritten as a 2-surface integral of an 
appropriate superpotential on the `spatial infinity' of $\Sigma$ (see also 
\cite{Shiromizu,NaShiMa,ShiIdTo}).

According to the traditional prescription \cite{AD}, the so-called
Abbott--Deser (AD) energy is ${\tt Q}[\bar K]$ in which $\bar K^a$ is the
`time translational Killing field', given explicitly e.g. by (\ref{eq:2.3}).
However, this Killing field is \emph{not} timelike on any neighbourhood of
some asymptotic end of $\Sigma$. (In fact, as we saw in subsection
\ref{sub-2.1.2}, this holds for \emph{any} Killing field in a neighbourhood
of $\mathscr{I}^+$.) Nevertheless, this can still be called `energy', but its
interpretation is not so obvious. For alternative forms of this energy
expression see \cite{ABK1} and (under additional conditions)
\cite{ShiIdTo,Luoetal}.

Although in certain special cases the positivity of the AD energy can be
proven \cite{Shiromizu,ShiIdTo,Luoetal}, in general it can have any sign
\cite{NaShiMa,LiZh}. Indeed, in the light of the results of \cite{ABK2} on
the energy flux through $\mathscr{I}^+$ of the linear gravitational
perturbations preserving the intrinsic conformal flatness of $\mathscr{I}^+$
(see subsection \ref{sub-2.2.1}), this notion of energy does not seem to
have the rigidity property. To resolve these and above conceptual
difficulties, it has already been suggested to use the \emph{conformal
Killing vector} $K^a_0$, given in (\ref{eq:2.4}), as the generator of the
total de Sitter energy \cite{ShiIdTo}. This conformal Killing field is used
in the next subsection, too.

\subsection{A spinorial expression of Kastor and Traschen}
\label{sub-3.2}

Using the original Dirac spinor form of the Nester--Witten form
\cite{Witt,Ne}, the Witten-type energy positivity proof of
\cite{Gibbonsetal}, originally given for $\Lambda<0$, has been successfully
adapted to the $\Lambda>0$ case and yielded a positivity argument by Kastor
and Traschen in \cite{KaTr} (see also \cite{ShiIdTo}). Here, the
energy-momentum tensor of the matter fields is assumed to satisfy the
dominant energy condition, and, as the boundary condition for the (modified)
Witten equation, the Dirac spinor constituents $\Psi$ of the \emph{conformal}
Killing vectors of the background de Sitter spacetime are used.

A generalization $C^a$ of the conserved current $\bar K_bt^{ba}$ of subsection 
\ref{sub-3.1}, built from $t_{ab}$, the `perturbation' $\gamma_{ab}$ and an 
\emph{arbitrary} vector field $\xi^a$, is also given in \cite{KaTr}: this 
$C^a$ is $\bar\nabla_a$-divergence free, and if $\xi^a$ is chosen to be a 
Killing vector of the de Sitter background, then it reduces to $\bar K_a
t^{ab}$. Then, it is shown that the flux integral of $C^b$ on $\Sigma$ with 
the \emph{conformal Killing vector} $\xi^a$ determined by the Dirac spinor 
$\Psi$ coincides with the spinorial construction above. Therefore, the 
present spinorial expression of these authors is a non-trivial 
\emph{generalization} of the original construction in \cite{AD}.

\subsection{Asymptotically-Schwarzschild/Kerr--de Sitter data: work 
of Chru\'sciel and collaborators}
\label{sub-3.3}

Chru\'sciel with a range of collaborators has been investigating definitions
of mass in a variety of space-times with $\Lambda$ positive, negative or zero
for many years. The investigations are typically motivated by adherence to
a Hamiltonian approach to space-times, following the monograph \cite{cjk1},
analysing Cauchy data to arrive at definitions of mass and momentum. In this
subsection we review their work on asymptotically-Schwarzschild--de Sitter
data and asymptotically-Kerr--de Sitter data. These are discussed in
\cite{cjk3} and \cite{cjk4} respectively and these in turn rely on families
of initial data for the Einstein equations shown to exist in \cite{cpo} and
\cite{cort} respectively.

The Schwarzschild--de Sitter metric, sometimes called the Kottler solution,
can be defined in space-time dimension $n+1$ as

\begin{equation}\label{ch1}
ds^2=-Vdt^2+\frac{dr^2}{V}+r^2\tilde{h},
\end{equation}
where $\tilde{h}$ is the metric of the standard (round) unit $(n-1)$-sphere
and

\begin{equation}\label{ch2}
V=1-\frac{2m}{r^{n-2}}-\frac{r^2}{\ell^2},
\end{equation}
with $m,\ell$ real positive constants. It is straightforward to check that
the metric (\ref{ch1}) satisfies the Einstein equations with cosmological
constant\footnote{Here the Einstein equations are taken to be $G_{ab}+
\Lambda g_{ab}=0$} $\Lambda=n(n-1)/(2\ell^2)$.

Typically one wants $\partial/\partial t$ to be a time-like Killing vector
at least somewhere, so that $V$ must be positive for some $r$, and it will
be, with two positive simple zeroes, provided

\begin{equation}\label{ch2a}
m^2\Lambda^{n-2}<c_n:=\frac{1}{n^2}\left(\frac{(n-1)(n-2)}{2}\right)^{n-2}.
\end{equation}

On a surface of constant $t$ the metric (\ref{ch1}) defines the spatial
metric

\begin{equation}\label{ch3}
g=\frac{dr^2}{V}+r^2\tilde{h},
\end{equation}
which is conformally-flat (by spherical symmetry) with constant scalar
curvature $R=n(n-1)/\ell^2$ and necessarily satisfies the Einstein constraint
equations with vanishing second fundamental form (so this is
\emph{time-symmetric} data). Given (\ref{ch2a}), the function $V$ has two
real distinct positive zeroes in $r$ as noted, but the metric can be extended
through both of these (in the style of the Kruskal extension) to obtain a
complete, spherically-symmetric, conformally-flat metric with constant scalar
curvature. Such metrics are called \emph{Delaunay metrics} in \cite{cpo} by
analogy with Delaunay surfaces which are complete, rotationally-symmetric,
constant mean curvature surfaces in $\mathbb{R}^3$. The Delaunay metrics
also arise from singular solutions of the Yamabe problem on $S^n$ (see the
discussion in \cite{cpo}).

A similar but more general class of solutions of the Einstein equations than
those in (\ref{ch1}) was published by Birmingham \cite{birm}. The metric
looks the same but $\tilde{h}$ is taken to be a Riemannian Einstein metric
on a compact $(n-1)$-manifold with scalar curvature $\tilde{R}$, and $V$ is
changed to

\begin{equation*}
V=\frac{\tilde{R}}{(n-1)(n-2)}-\frac{2m}{r}-\frac{r^2}{\ell^2}.
\end{equation*}
With positive $m$, we need positive $\tilde{R}$ to have $V$ anywhere positive
and then constant rescaling and redefinition of $r$ can be used to set
$\tilde{R}=(n-1)(n-2)$ to recover the previous expression for $V$. Now one
can introduce a notion of \emph{generalised Delaunay metrics} with this
$\tilde{h}$ in place of the previous $\tilde{h}$.

In \cite{cpo} the authors use gluing techniques to produce new time-symmetric
data for the Einstein equations which have many \emph{exactly} Delaunay or
generalised Delaunay ends, which is to say that the ends are exactly Delaunay
or generalised Delaunay for infinite stretches. These ends can then in turn
be truncated and glued on to other regions to produce data on connected sums
or wormholes. In \cite{cjk3}, the authors use the Hamiltonian method to assign
masses to the simplest case of initial data in this class, a single region
with several Delaunay ends. The result is that the Hamiltonian mass of each
end is precisely the mass parameter $m$ in the appropriate metric (\ref{ch3})
(and \emph{a fortiori} it is positive).

In \cite{cort}, Cortier generalised the work in \cite{cpo} to produce initial
data which are deformations of that for Kerr--de Sitter (but only in
space-time dimension 4, so the data surface has dimension 3). Again one has
an infinite periodic metric but now with a particular nonzero periodic second
fundamental form as well. Then in \cite{cjk4}, the authors use the Hamiltonian
method to assign both mass and angular momentum to these data.

This work, in associating a mass with each Delaunay or generalised Delaunay
end, is reminiscent of the calculation of Penrose's quasi-local mass for
various 2-spheres in conformally-flat, time-symmetric initial data for the
vacuum Einstein equations (so $\Lambda=0$) in \cite{kpt}. In particular, the
mass defined can be thought of as a `mass at spatial infinity', given a
Delaunay or generalised Delaunay end that extends to spatial infinity, but
it is not a mass at null infinity.


\section{TBS type total energy-momenta}
\label{sec-4}

In contrast to the ADM type energy-momenta, the TBS type expressions are
associated with \emph{closed orientable 2-surfaces} ${\cal S}$, that is
`cuts', rather than points of $\mathscr{I}^+$. Then, to be potentially able
to prove a formula analogous to Bondi's mass-loss for the TBS type
expressions, we should have, in fact, a \emph{foliation} of $\mathscr{I}^+$
by such cuts. Thus, the spacelike hypersurfaces $\Sigma$ that intersect
$\mathscr{I}^+$ in the given cuts should be analogous to the hypersurfaces
$\Sigma_{\hat\tau}$ discussed at the end of subsection \ref{sub-2.1.1}.

\subsection{Linearization around de Sitter}
\label{sub-4.1}

In \cite{Tod15} some general twistor theory of de Sitter spacetime is
recorded, sufficient to describe Penrose's quasi-local mass construction at
any space-like 2-sphere for linear gravitational perturbations of de Sitter.
This includes 2-spheres at $\mathscr{I}^+$, and it is pointed out that the
absence of timelike Killing fields near $\mathscr{I}^+$ entails that this
quasi-local mass does not have a positivity property, in contrast to the
$\Lambda=0$ and $\Lambda<0$ cases. It can be the basis of a definition in
asymptotically-de Sitter space-times but it won't have positivity or
rigidity properties.

\subsection{Three Suggestions of Penrose}
\label{sub-4.2}

In \cite{Pe11} Penrose considered the problem of defining a cosmological
total mass in the presence of a positive cosmological constant. He was
motivated, at least in part, by his Conformal Cyclic Cosmology (or CCC)
rather than by a consideration of isolated systems, which was the motivation
of Ashtekar and colleagues considered above. Thus Penrose sought a definition
at $\mathscr{I}^+$ of one aeon which was suitable for carrying through to
the next aeon. Interestingly, Penrose considered but rejected the idea of
working in a Poincar\'e patch and using the  Killing vector of the
steady-state universe to define energy (section 3 of \cite{Pe11}) because
galaxies beyond the cosmological event horizon, where this Killing vector is
space-like, would be represented as having super-luminal velocities. This
could be a problem in a cosmological setting but one that Ashtekar \emph{et.
al.} avoid by considering only isolated bodies.

\medskip

Penrose makes three concrete suggestions:

(1) To seek an expression motivated by the conserved currents that one
constructs from a trace-free energy-momentum tensor and a conformal Killing
vector of de Sitter space. In CCC one expects matter to become massless
near $\mathscr{I}^+$, so that $T_{ab}$ will become trace-free, and in de
Sitter space there do exist time-like \emph{conformal} Killing vectors. This
suggestion is not worked out in detail but the insight also underlay the
earlier work of \cite{KaTr}, \cite{Shiromizu} and \cite{ShiIdTo}.

(2) To use his quasi-local energy-momentum construction, as discussed above
(see subsection \ref{sub-4.1}), at $\mathscr{I}^+$. This construction is
defined for a space-like 2-surface ${\cal S}$, usually a topological sphere.
One first solves an elliptic system for the 2-surface twistors on ${\cal S}$,
which are particular 2-component spinor fields, then forms an integrand
linear in the spacetime curvature but with the cosmological constant term
removed and integrates over ${\cal S}$. This is a well-defined procedure but
the result does not have any positivity or therefore rigidity property. In
particular it can be zero even in the presence of non-trivial curvature
\cite{SzT}: applied to 2-spheres on the $\mathscr{I}^+$ of the
Schwarzschild--de Sitter space-time it gives zero if the 2-sphere ${\cal S}$
is homologous to a point on $\mathscr{I}^+$ and a non-zero constant but of
either sign if ${\cal S}$ surrounds the source.

(3) To use the original TBS energy expression, given in the Newman--Penrose
form as

\begin{equation}
E=\frac{2}{\kappa}\oint_{S}\bigl(\sigma\sigma'-\psi_2\bigr){\rm d}{\cal S},
\label{eq:4.1}
\end{equation}
where $\psi_2$ is the $o^Ao^B\iota^C\iota^D$ component of the Weyl curvature
spinor and $\sigma$ and $\sigma'$ are the asymptotic shear of the two null
geodesic congruences hitting the cut ${\cal S}$ of $\mathscr{I}^+$
orthogonally. This is the basis of the suggestion taken up by Saw
\cite{Saw1,Saw2,Saw6}, following Frauendiener \cite{Joerg97,Joerg}.

\subsection{Mass-loss and the solution of the NP equations}
\label{sub-4.3}

The key property of the TBS mass in the asymptotically flat context is 
mass-loss: the mass, as a function of the retarded time coordinate, is a 
monotonically \emph{decreasing} function. Thus, accepting (\ref{eq:4.1}) 
as the definition of the TBS energy in the $\Lambda>0$ case, the natural 
question is whether or not the analogue of the mass-loss formula can be 
derived. Such a formula could be based directly on the analysis of the 
Bianchi identities, or on the general integral formula of Frauendiener 
\cite{Joerg97,Joerg}. However, to derive this, one should solve the 
Newman--Penrose spin coefficient equations to an appropriately high order. 
This solution, both for the vacuum and electro-vacuum in the physical 
spacetime, is given by Saw in \cite{Saw1} and \cite{Saw2}, respectively 
(see also \cite{Saw3}). The analogue of the mass-loss formula is derived 
from the Bianchi identity in \cite{Saw1,Saw2}, using an integral identity 
in \cite{Saw6}. In \cite{Saw6}, with $E$ as in (\ref{eq:4.1}),  he finds 

\begin{equation}
\dot E=-\frac{2}{\kappa}\oint_{\cal S}\Bigl(\vert\dot\sigma\vert^2+
\frac{\Lambda}{2}\vert\edth^\prime\sigma\vert^2 -\frac{\Lambda}{6}\vert
\edth\sigma\vert^2 +\frac{2\Lambda}{9}\vert\sigma
\vert^4+\frac{\Lambda^2}{18}\Re(\bar\sigma\psi_0)\Bigr){\rm d}{\cal S},
\label{eq:4.2}
\end{equation}
where the dot denotes derivative with respect to the parameter $u$ which 
labels the actual cut ${\cal S}$ of the foliation of $\mathscr{I}^+$ by 
2-surfaces ${\cal S}_u$, $\Re$ denotes `real part', ${\edth}$ and ${\edth}
^\prime$ are the standard GHP edth and edth-prime operators, respectively, and 
$\psi_0$ is the $o^Ao^Bo^Co^D$ component of the Weyl curvature spinor. Saw 
shows that, in the absence of incoming radiation, so that $\psi_0=0$, and for 
purely quadrupole gravitational radiation (which here is taken to mean that 
$\edth\sigma=0$, which would imply, for a sphere with constant Gauss 
curvature, that $\sigma$ is proportional to a combination of the ${}_2Y_{2m}$ 
spin weighted spherical harmonics) $\dot E$ is nonpositive and vanishes only 
for zero $\sigma$. He also notes, in \cite{Saw1}, that the vanishing of 
$\sigma$ on $\mathscr{I}^+$ is equivalent to the intrinsic conformal flatness 
of $\mathscr{I}^+$, in agreement with \cite{ABK1}. 

Saw also makes use of another expression for $\dot{E}$, equivalent to 
(\ref{eq:4.2}) by an identity for functions on a sphere, namely 

\begin{equation}
\dot E=-\frac{2\Lambda}{3\kappa}\oint_{\cal S}K\vert\sigma\vert^2{\rm d}
{\cal S}-\frac{2}{\kappa}\oint_{\cal S}\Bigl(\vert\dot\sigma\vert^2+
\frac{\Lambda}{3}\vert\edth^\prime\sigma\vert^2+\frac{2\Lambda}{9}\vert\sigma
\vert^4+\frac{\Lambda^2}{18}\Re(\bar\sigma\psi_0)\Bigr){\rm d}{\cal S},
\label{eq:4.2a}
\end{equation}
where $K$ is the Gauss curvature of ${\cal S}$ (which need not have fixed 
sign). This is the basis of a suggestion he makes for a different analogue of 
the TBS mass in the $\Lambda>0$ case, namely 

\begin{equation*}
E_\Lambda(u):=E(u)+\frac{2\Lambda}{3\kappa}\int^u_{u_0}\Bigl(\oint_{{\cal S}
_{\bar u}}K\vert\sigma\vert^2{\rm d}{\cal S}_{\bar u}\Bigr){\rm d}\bar u
\end{equation*}
with $E(u)$ as in (\ref{eq:4.1}), and for some $u_0$. Now the $u$-derivative 
of $E_\Lambda$ consists of just the second integral in (\ref{eq:4.2a}) and in 
the absence of incoming radiation, but this time without the restriction to 
quadrupole radiation, it is therefore \emph{negative} (strictly speaking it 
is nonnegative and vanishes only for zero $\sigma$). In this mass-loss formula 
there is a term proportional to $\vert\dot\sigma\vert^2$ that is familiar 
from the case of $\Lambda=0$, but there are also terms involving the shear 
and its angular derivatives but undotted. However, this proposed new 
expression depends not only on the instantaneous state of the physical system 
at $u$, but on the whole history of the system until $u$. A summary of the 
results of the investigations of Saw prior to \cite{Saw6} is given in 
\cite{Saw4,Saw5}.

\subsection{A general TBS-type spinorial expression}
\label{sub-4.4}

As noted above, a key property of the TBS energy in the asymptotically flat
case is its positivity and rigidity. Since in the $\Lambda=0$ and $\Lambda<0$
cases the total energy-momenta could be recovered from a unified form based
on the integral of the Nester--Witten form (\ref{eq:1.2}), and moreover the
simplest energy positivity proof is arguably the one based on the use of
2-component spinors and Witten-type arguments \cite{HT,Gibbonsetal},
it seems natural to try to formulate the TBS energy-momentum in the presence
of a positive $\Lambda$ in this framework, too. This was done in \cite{SzT}.
However, since it is not \emph{a priori} clear what the `asymptotic time
translations' near $\mathscr{I}^+$ should be, in the investigations of
\cite{SzT} the spinor fields at the cut are initially left unspecified.
Surprisingly enough, the requirement of the finiteness of the resulting
integral together with the desire to have a Witten-type energy positivity
proof determine the spinor fields in (\ref{eq:1.2}) at the cut: \emph{They
should solve Penrose's 2-surface twistor equations.} Thus, \emph{the boundary
conditions} for the Witten spinors, i.e. the spinor constituents of what one
could considered to be the `asymptotic translations', \emph{come out of the
formalism}.

The integral of the Nester--Witten form with these boundary conditions
defines a $4\times4$ complex Hermitian matrix with the structure

\begin{equation*}
H=\left(\begin{array}{cc}
      {\tt P} & {\tt Q} \\
 -\bar{\tt Q} & \bar{\tt P} \\
      \end{array}\right),
\end{equation*}
where ${\tt P}$ is a $2\times2$ Hermitian, and ${\tt Q}$ is a complex
anti-symmetric matrix. If the matter fields satisfy the dominant energy
condition, then it has been shown that $H$ is \emph{positive definite}
(`positivity'), and it is vanishing precisely when the past domain of
dependence of the (asymptotically hyperboloidal) spacelike hypersurface
$\Sigma$ for which $\Sigma\cap\mathscr{I}^+$ is just the given cut ${\cal S}$
is locally isometric to the de Sitter spacetime (`rigidity'). The symmetry
group of the 2-surface twistor space $\mathbb{T}$ on ${\cal S}$ is $(0,
\infty)$ times the spin group of $SO(1,5)$. If there exists a volume 4-form
on $\mathbb{T}$, then the TBS-type mass can be defined as the determinant of
$H$. For example, if there is a scalar product on $\mathbb{T}$ (which with
$(+,+,-,-)$ signature is guaranteed e.g. when $\mathscr{I}^+$ is intrinsically
conformally flat), then such a volume form exists. In these special cases the
symmetry group reduces to the spin group of $SO(1,5)$ and to the spin group
of the de Sitter group $SO(1,4)$, respectively.

The same general analysis can be repeated in the $\Lambda<0$ and $\Lambda=0$
cases, too \cite{SzT2}. If $\Lambda<0$, then the result is similar but the
symmetry group of $\mathbb{T}$ is $(0,\infty)$ times the spin group of
$SO(3,3)$, which, in the presence of the reflective boundary condition,
reduces to the spin group of the anti-de Sitter group $SO(2,3)$. If $\Lambda
=0$, then $\mathbb{T}$ splits to the direct sum $\mathbb{T}_0\oplus
\mathbb{T}_0$, where $\mathbb{T}_0$ is the space of the spinor constituents
of the BMS translations, ${\tt Q}$ in $H$ is zero, ${\tt P}$ reduces to the
TBS 4-momentum, and the symmetry group of $\mathbb{T}$ is $SL(2,\mathbb{C})
\times SL(2,\mathbb{C})$. Therefore, the above construction is a natural
generalization of the TBS 4-momentum of asymptotically flat spacetimes to
the $\Lambda>0$ case. It has positivity and rigidity, at the cost of being a
matrix with six independent real components rather than a Lorentz scalar or
vector.

\subsection{Characteristic data: the work of Chru\'sciel--Ifsits }
\label{sub-4.5}

In \cite{ChrusIf} Chru\'sciel and Ifsits define a mass from data for the
Einstein equations given on an outgoing null hypersurface $\mathcal{N}$,
which could be the null cone of a point, in a conformally-compactifiable
$(n+1)$-dimensional space-time. The main interest in \cite{ChrusIf} is with
$\Lambda>0$ but all values of $\Lambda$ are allowed, and the method is to
construct a Bondi coordinate system at $\mathcal{N}$ and solve for the
space-time metric and connection at $\mathcal{N}$ in terms of free data,
making assumptions about asymptotic decay rates, for example of any matter
content, as required. The paper proposes a definition of mass from a
consideration of the TBS mass defined for $\Lambda=0$. The generalised
definition is checked against the $\Lambda=0$ case, for which a corresponding
calculation was presented in \cite{cp}, and, by consideration of a variety
of examples, with the $\Lambda<0$ case, which is fairly well understood.

The calculation leading to the definition of mass naturally leads to the
definition of a \emph{renormalised volume} for $\mathcal{N}$. This is
defined from the integral $V(r)$ of the area $A(r)$ of sections\footnote{In
space-time dimension $n+1$, $A$ is the $(n-1)$-dimensional volume of these
sections.} of $\mathcal{N}$ of constant $r$ where $r$ is an affine parameter
along the generators of $\mathcal{N}$. Given an origin for $r$, which would
be the vertex of $\mathcal{N}$ if $\mathcal{N}$ were a light-cone but other
choices are possible, the authors obtain an expansion:

\begin{equation*}
V(r)=\int_{0}^rA(r')dr'=a_3r^3+a_2r^2+a_1r+a_\ell\log r+a_0+a_{-1}r^{-1}+o(r^{-1}),
\end{equation*}
for coefficients $a_k$ given in terms of the data and quantities obtained
from the data. Then the coefficient $a_0=V_{ren}$ is the renormalised volume.

In discussion of the result in the last section of the paper it is observed
that the mass defined is geometric and gauge-invariant and coincides in
cases of $\Lambda\leq 0$ with other accepted definitions, but that it is not
obviously non-negative in general, nor \emph{rigid} in the sense that
vanishing mass implies that the space-time is exactly de Sitter inside the
cone $\mathcal{N}$.


\section{Total mass in closed universes}
\label{sec-5}

The construction of the total mass for closed universes, introduced first
for $\Lambda=0$ (and mentioned at the end of subsection \ref{sub-1.1.3}),
can be generalized in a straightforward way to closed universes with
positive $\Lambda$ \cite{Sz13}. The basis of this construction is the
observation that, in the Witten type gauge, the $\lambda^A\bar\lambda^{A'}$
-component of the hypersurface integral form of the spinorial expression of
\emph{all} the energy-momentum expressions above (independently of the sign
of $\Lambda$) takes the manifestly positive definite expression

\begin{equation*}
{\tt P}(\lambda):=\frac{\sqrt{2}}{\kappa}\Vert{\cal D}_{(AB}\lambda_{C)}
\Vert^2_{L_2}+\int_\Sigma t^aT_{ab}\lambda^B\bar\lambda^{B'}{\rm d}\Sigma.
\end{equation*}
Here $\Sigma$ is the asymptotically flat/hyperboloidal hypersurface whose
boundary `at infinity' is just the 2-surface ${\cal S}$, the $L_2$-norm is
defined on this $\Sigma$ with $\sqrt{2}t_{AA'}$ as the pointwise positive
definite Hermitian scalar product on the spinor spaces, and ${\cal D}_{AB}$
is the unitary spinor form of the Sen derivative operator on $\Sigma$ (see
\cite{Tod84}).

However, if $\Sigma$ is a \emph{compact} Cauchy surface in a closed universe,
e.g. when $\Sigma$ is analogous to the leaves $\Sigma_t$ of the global
foliation of the de Sitter spacetime given in subsection \ref{sub-2.1.1},
then the above quantity can be formed even when $T_{ab}$ is replaced by $T
_{ab}+g_{ab}\Lambda/\kappa$ with positive $\Lambda$ and even for any spinor
field $\lambda^A$ with the normalization $\Vert\lambda\Vert^2_{L_2}=\sqrt{2}$.
The total mass ${\tt M}$ (in fact, mass density) associated with $\Sigma$ is
defined to be just the \emph{infimum} of ${\tt P}(\lambda)$ on the set of
spinor fields satisfying the normalization above. Then, the spinor fields
$\lambda^A$ for which ${\tt P}(\lambda)={\tt M}$ holds are precisely the
eigenspinors in the eigenvalue equation ${\cal D}^{AA'}{\cal D}_{A'B}\lambda
^B=(3\kappa/8){\tt M}\lambda^A$. Thus, the mass ${\tt M}$ could have been
defined as \emph{the first eigenvalue} in this eigenvalue problem. Clearly,
$\kappa{\tt M}\geq\Lambda$ by construction, but it has the non-trivial
\emph{rigidity property}: $\kappa{\tt M}=\Lambda$ if and only if the whole
spacetime is locally isometric to the de Sitter spacetime and the Cauchy
surface $\Sigma$ is homeomorphic to $S^3/G$, where $G$ is a discrete
subgroup of $SU(2)\approx S^3$.

\bigskip
\bigskip
\noindent
The authors are grateful to Gy\"orgy Wolf for drawing the figures. 



\begin{thebibliography}{99.}%


\bibitem{Trautman} A. Trautman, Conservation laws in general relativity, in
             {\it Gravitation: An Introduction to Current Research}, Ed: L.
             Witten, pp. 169-198, Wiley, New York 1962

             A. Trautman, F. A. E. Pirani, H. Bondi, Lectures on General
             Relativity, Brandeis Summer Institute in Theoretical Physics,
             New Jersey 1964
\bibitem{MTW} C. Misner, K. Thorne, J. A. Wheeler, {\it Gravitation},
             Freeman, San Francisco 1973
\bibitem{Goldberg} J. Goldberg, Invariant transformations, conservation laws
            and energy-momentum, in {\it General Relativity and Gravitation},
            Ed.: A. Held, vol 1, pp 469-489, Plenum, New York 1980

\bibitem{Sparl82} G. A. J. Sparling, Twistors, spinors and the Einstein
             vacuum equations, Pittsburgh preprint, 1982; also in {\it
             Further Advances in Twistor Theory, III: Curved Twistor Spaces},
             Ed.: L. Mason {\it etal}, pp. 179-186, Chapman and Hill,
             London 2001
\bibitem{DuMa} M. Dubois-Violette, J. Madore, Conservation laws and
             integrability conditions for gravitational and Yang-Mills
             equations, Commun. Math. Phys. {\bf 108} 213-223 (1987)
\bibitem{Joerg89} J. Frauendiener, Geometric description of energy-momentum
             pseudotensors, Class. Quantum Grav. {\bf 6} L237-L241 (1989)
\bibitem{Sz92} L. B. Szabados, On canonical pseudotensors, Sparling's form
              and Noether currents, Class. Quantum Grav. {\bf 9} 2521-2541
              (1992); preprint KFKI-1991-29/B

\bibitem{SzRev} L. B. Szabados, Quasi-local energy-momentum and angular
              momentum in GR, Living Rev. Relativity {\bf 12} (2009) No 4.


\bibitem{HE} S.W. Hawking, G. F. R. Ellis, {\it The Large Scale Structure of
             Spacetime}, Cambridge University Press, Cambridge 1973

\bibitem{ADM} R. Arnowitt, S. Deser, C. W. Misner, The dynamics of general
              relativity, in {\it Gravitation: An Introduction to Current
              Research}, pp. 227--265, Ed.: Witten, L., Wiley, New York,
              London, 1962, arXiv: gr-qc/0405109
\bibitem{Trautman58} A. Trautman, Lectures on general relativity (Lectures
             at King's College in London, May-June 1958), Gen. Relativity
             Grav. {\bf 34} 721-762  (2002)

\bibitem{Bondietal} H. Bondi, Gravitational waves in general relativity,
              Nature, {\bf 186} 535 (1960)

              H. Bondi, M. G. J. van der Burg, A. W. K. Metzner,
              Gravitational waves in general relativity. VII. Waves from
              axi-symmetric isolated systems, Proc. R. Soc. London, Ser. A
              {\bf 269} 21-52 (1962)
\bibitem{Sachs} R. K. Sachs, Asymptotic symmetries in gravitational theory,
              Phys. Rev. {\bf 128} 2851--2864 (1962)

\bibitem{Pe65} R. Penrose, Zero rest-mass fields including gravitation:
             asymptotic behaviour, Proc. Roy. Soc. (London) A {\bf 284}
             159-203 (1965)
\bibitem{PR2} R. Penrose, W. Rindler, {\it Spinors and Spacetime}, vol 2,
              Cambridge University Press, Cambridge 1986

\bibitem{Newm} R. P. A. C. Newman, The global structure of simple space-times,
              Commun. Math. Phys., {\bf{123}} 17--52 (1989)

\bibitem{Ge}  R. Geroch, Asymptotic structure of spacetime, in {\it
              Asymptotic structure of spacetime}, Ed. F. P. Esposito, L.
              Witten, Plenum Press, New York 1977
\bibitem{NT80} E. T. Newman, K. P. Tod, Asymptotically flat spacetimes, in
             {\it General Relativity and Gravitation}, vol 2, Ed. A. Held
             (New York, Plenum) 1980
\bibitem{JoergLivRev} J. Frauendiener, Conformal infinity, Living Rev.
             Relativ. (2004) 7: 1. https://doi.org/10.12942/lrr-2004-1

\bibitem{GeWi} R. Geroch, J. Winicour, Linkages in general relativity, J.
               Math. Phys. {\bf 22} 803-812 (1981)

\bibitem{ABK1} A. Ashtekar, B. Bonga, A. Kesavan, Asymptotics with a positive
            cosmological constant: I. Basic framework, Class. Quant. Grav.
            {\bf 32} 025004 (41pp) (2014), arXiv: 1409.3816 [gr-qc]

\bibitem{slowfalloff} L. Andersson, P. T. Chru\'sciel, Hyperboloidal Cauchy
             data for vacuum Einstein equations and obstructions to smoothness
             of null infinity, Phys. Rev. Lett. {\bf 70} 2829-2832 (1993),
             arXiv: gr-qc/9304019

             J. A. Valiente-Kroon, A new class of obstructions to the
             smoothness of null infinity, Commun. Math. Phys. {\bf 244}
             133-156 (2004)

\bibitem{Sz01} L. B. Szabados, On certain quasi-local spin-angular-momentum
              expressions for large spheres near the null infinity, Class.
              Quantum Grav. {\bf 18} 5487-5510 (2001),  arXiv: gr-qc/0109047

\bibitem{WiTa} J. Winicour, L. Tamburino, Lorentz-covariant gravitational 
              energy-momentum linkages, Phys. Rev. Lett. {\bf 15} 601-605 
              (1965) 

\bibitem{Ne} J. M. Nester, A new gravitational energy expression with a
             simple positivity proof, Phys. Lett. A {\bf 83} 241-241 (1981)
\bibitem{Witt} E. Witten, A new proof of the positive energy theorem,
              Commun. Math. Phys. {\bf 80} 381-402 (1981)

\bibitem{ScY} R. Schoen, S.-T. Yau, Proof of the positive mass theorem, II.
              Commun. Math. Phys. {\bf 79} 231-260 (1981)
\bibitem{HT} G. Horowitz, K. P. Tod, A relation between local and total
             energy in general relativity, Commun. Math. Phys. {\bf 85}
             429--447 (1982)
\bibitem{RT} O. Reula, K. P. Tod, Positivity of the Bondi energy, J. Math.
             Phys. {\bf 25} 1004-1008 (1984)
\bibitem{Gibbonsetal} G. W. Gibbons, S. W. Hawking, G. T. Horowitz, M. J.
             Perry, Positive mass theorems for black holes, Cummun. Math.
             Phys. {\bf 88} 295-308 (1983)

\bibitem{ChCh} Y. Choquet-Bruhat, D. Christodoulou, Elliptic systems in
             $H_{s.\delta}$ spaces on manifolds which are Euclidean at infinity,
             Acta. Math. {\bf 146} 129-150 (1981)
\bibitem{SzT} L. B. Szabados, P. Tod, A positive Bondi-type mass in
              asymptotically de Sitter spacetimes, Class. Quantum Grav.
              {\bf 32} (2015) 205011 (pp 51), arXiv: 1505.06637 [gr-qc]

\bibitem{Sz12} L. B. Szabados, Mass, gauge conditions and spectral properties
             of the Sen--Witten and 3-surface twistor operators in closed
             universes, Class. Quantum Grav. {\bf 29} 095001 (2012) (30pp),
             arXiv: 1112.2966 [gr-qc]
\bibitem{Sz13a} L. B. Szabados, On the total mass of closed universes,
           Gen. Rel. Grav. {\bf 45} 2325-2339 (2013), arXiv:1212.0147 [gr-qc]


\bibitem{AD} G. Abbott, S. Deser, Stability of gravity with a cosmological
             constant, Nucl. Phys. B {\bf 195} 76-96 (1982)

\bibitem{AM} A. Ashtekar, A. Magnon, Asymptotically anti-de Sitter spacetimes,
             Class. Quantum Grav. {\bf 1} L39-L44 (1984)
\bibitem{Ha83} S. W. Hawking, The boundary conditions for gauged
             supergravity, Phys. Lett. {\bf 126 B} 175-177 (1983)
\bibitem{Tod84} K. P. Tod, Three-surface twistors and conformal embedding,
             Gen. Rel. Grav. {\bf 16} 435-43 (1984)

\bibitem{Friedrich} H. Friedrich, G. Nagy, The initial boundary value problem
             for Einstein's vacuum field equation, Commun. Math. Phys. {\bf
             201} 619-655 (1999)

             H. Friedrich, Initial boundary value problems for
             Einstein's field equations and geometric uniqueness, Gen.
             Relativ. Grav. {\bf 41} 1947-1966 (2009),  arXiv: 0903.5160

\bibitem{Friedrich14a} H. Friedrich, On the AdS stability problem, Class.
             Quantum Grav. {\bf 31} (2014) 105001, arXiv: 1401.7172

\bibitem{ChruscielNagy} P. T. Chru\'sciel, G. Nagy, The mass of spacelike
             hypersurfaces in asymptotically anti-de Sitter space-times, Adv.
             Theor. Math. Phys. {\bf 5} 697-754 (2002), arXiv: gr-qc/0110014
\bibitem{Ke} R. Kelly, Asymptotically anti-de Sitter spacetimes, Twistor
             Newsletter, No 20, 11--23 (1985)

             R. Kelly, P. Tod, Penrose's quasi-local mass for asymptotically
             anti-de Sitter space-times, arXiv: 1505.00214



\bibitem{Riessetal} A. G. Riess {\it etal}, Observational evidence from
             supernovae for an accelerating Universe and a cosmological
             constant, Astron. J. {\bf 116} 1009-1038 (1998), arXiv:
             astro-ph/9805201
\bibitem{Perlmutteretal} S. Perlmutter {\it etal}, Measurements of Omega and
             Lambda from 42 high-redshift supernovae, Astrophys. J. {\bf 517}
             565-586 (1999), arXiv: astro-ph/9812133

\bibitem{Pe10} R. Penrose, {\it Cycles of Time}, The Bodley Head, London 2010

\bibitem{Pe11} R. Penrose, On cosmological mass with positive $\Lambda$,
               Gen. Relat. Grav. {\bf 43} 3355-3366 (2011)


\bibitem{Tod15} P. Tod, Some geometry of de Sitter space, arXiv: 1505.06123
             [gr-qc]

\bibitem{Luoetal} M. Luo, N. Xie, X. Zhang, Positive mass theorems for
             asymptotically de Sitter spacetimes, Nucl. Phys. B {\bf 825}
             98-118 (2010), arXiv: 0712.4113v3 [math.DG]

\bibitem{KaTr} D. Kastor, J. Traschen, A positive energy theorem for
             asymptotically de Sitter spacetimes, Class. Quantum Grav.
             {\bf 19} 5901-5920 (2002)

\bibitem{ABK2} A. Ashtekar, B. Bonga, A. Kesavan, Asymptotics with a positive
            cosmological constant: II. Linear fields on de Sitter spacetime,
            Phys. Rev. D {\bf 92} 044011 (2015), arXiv: 1506.06152 [gr-qc]
\bibitem{ABK3} A. Ashtekar, B. Bonga, A. Kesavan, Asymptotics with a positive
            cosmological constant: III. The quadruple formula, Phys. Rev. D
            {\bf 92} 10432 (2015), arXiv: 1510.05593 [gr-qc]
\bibitem{ABK4} A. Ashtekar, B. Bonga, A. Kesavan, Gravitational waves from
            isolated systems: Surprising consequences of a positive
            cosmological constant, Phys. Rev. Lett. {\bf 116} 051101 (2016),
            arXiv: 1510.04990 [gr-qc]
\bibitem{Ash17} A. Ashtekar, Implications of a positive cosmological constant
            for general relativity, Rep. Prog. Phys. {\bf 80} 102901 (2017),
            arXiv: 1706.07482 [gr-qc]


\bibitem{Friedrich14} H. Friedrich, Geometric asymptotics and beyond, in {\it
             Surveys in Differential Geometry}, Vol.20. Ed.: L. Bieri, S.-T.
             Yau, International Press, Boston 2015; arXiv: 1411.3854

             H. Friedrich, Smooth non-zero rest-mass evolution across timelike
             infinity, arXiv: 1311.0700 [gr-qc]


\bibitem{Shiromizu} T. Shiromizu, Positivity of gravitational mass in
             asymptotically de Sitter spacetimes, Phys. Rev. D {\bf 49}
             5026-5029 (1994)
\bibitem{NaShiMa} K. Nakao, T. Shiromizu, K. Maeda, Gravitational mass in
             asymptotically de Sitter spacetimes, Class. Quantum Grav.
             {\bf 11} 2059-2071 (1994)

\bibitem{ShiIdTo} T. Shiromizu, D. Ida, T. Torii, Gravitational energy,
             dS/CFT correspondence and cosmic no-hair, JHEP {\bf 0111}
             (2001) 010, arXiv: hep-th/0109057

\bibitem{LiZh} Z. Liang, X. Zhang, Spacelike hypersurfaces with negative
             total energy in de Sitter spacetime, J. Math. Phys. {\bf 53}
             022502 (2012), arXiv: 1105.1213v2 [gr-qc]


\bibitem{cjk1}  P. T. Chru\'sciel, J. Jezierski, J. Kijowski, {\it Hamiltonian
            field theory in the radiating regime}. Lecture Notes in Physics.
            Monographs, {\bf 70} Springer-Verlag, Berlin, 2002.
\bibitem{cjk3} P. T. Chru\'sciel, J. Jezierski, J. Kijowski, The Hamiltonian
           mass of asymptotically Schwarzschild-de Sitter space-times,
           Phys. Rev. {\bf{D 87}} (2013) 124015, arXiv: 1305.1014
\bibitem{cjk4} P. T. Chru\'sciel, J. Jezierski, J. Kijowski, Hamiltonian
          dynamics in the space of asymptotically Kerr-de Sitter spacetimes,
          Phys. Rev. {\bf{D 92}} (2015) 084030, arXiv: 1507.03868
\bibitem{cpo} P. T. Chru\'sciel, D. Pollack, Singular Yamabe metrics and
          initial data with exactly Kottler-Schwarzschild-de Sitter ends,
          Ann. Henri Poincar\'e {\bf{9}} (2008) 639-654, arXiv: 0710.3365
\bibitem{cort}  J. Cortier, Gluing construction of initial data with
          Kerr--de Sitter ends, Ann. Henri Poincar\'e {\bf{14}} (2013)
          1109--1134, arXiv: 1202.3688
\bibitem{birm} D. Birmingham, Topological black holes in anti-de Sitter
          space, Class. Quantum Grav. {\bf{16}} (1999) 1197--1205,  arXiv:
          hep-th/9808032

\bibitem{kpt}  K. P. Tod, Some examples of Penrose's quasi-local mass
          construction, Proc. Roy. Soc. London {\bf A388} 457--477 (1983)


\bibitem{Joerg97} J. Frauendiener, On an integral formula on hypersurfaces
             in general relativity, Class. Quantum Grav. {\bf 14} 2413-2423
             (1997)
\bibitem{Joerg} J. Frauendiener, Talk given at the workshop {\it Mathematics
             of CCC: Mathematical Physics with Positive Lambda}, The Clay
             Mathematics Institute, University of Oxford, September 11-13,
             2013

\bibitem{Saw1} V.-L. Saw, Mass-loss of an isolated gravitating system due
            to energy carried away by gravitational waves, with cosmological
            constant, Phys. Rev. D {\bf 94} 104004 (2016), arXiv: 1605.05151
            [gr-qc]
\bibitem{Saw2} V.-L. Saw, Behaviour of asymptotically electro-$\Lambda$
            spacetimes, Phys. Rev. D {\bf 95} 084038 (2017), arXiv:
            1608.06886 [gr-qc]
\bibitem{Saw3} V.-L. Saw, Peeling property as a consequence of the
            cosmological constant, arXiv: 1705.00435 [gr-qc]
\bibitem{Saw4} V.-L. Saw, Asymptotically simple spacetimes and mass loss due
            to gravitational waves, Int. J. Mod. Phys. D {\bf 26} (2017)
            1730027, arXiv: 1706.00160 [gr-qc]
\bibitem{Saw5} V.-L. Saw, Mass-loss due to gravitational waves with
            $\Lambda>0$, Mod. Phys. Lett. A {\bf 32} 1730020 (2017), arXiv:
            1704.07514 [gr-qc]
\bibitem{Saw6} V.-L. Saw, Bondi mass with a cosmological constant, Phys. Rev.
            D {\bf 97} 084017 (2018), arXiv: 1711.01808 [gr-qc]

\bibitem{SzT2} L. B. Szabados, P. Tod, (unpublished)


\bibitem{ChrusIf} P. T. Chru\'sciel, L. Ifsits, The cosmological constant and
            the energy of gravitational radiation, Phys. Rev. D {\bf 93}
            124075 (2016), arXiv: 1603.07018 [gr-qc]
\bibitem{cp} P. T. Chru\'sciel, T.-T. Paetz, The mass of light-cones, Class.
             Quantum Grav. {\bf{31}} (2014) 102001,  arXiv:1401.3789


\bibitem{Sz13} L. B. Szabados, On the total mass of closed universes with a
             positive cosmological constant, Class. Quantum Grav. {\bf 30}
             165013 (2013), arXiv: 1306.3863 [gr-qc]

\end{thebibliography}
\end{document}